%% file: main.tex
\documentclass[twocolumn]{article}

\usepackage[utf8]{inputenc}
\usepackage[T1]{fontenc}
\usepackage[margin=1.25in]{geometry}
\usepackage{authblk}
\usepackage{setspace}
\usepackage{graphicx}
\graphicspath{{./figures/}}
\usepackage{subcaption}
\usepackage{amsmath,amssymb}
\allowdisplaybreaks
\usepackage{booktabs}
\usepackage{multirow}
\usepackage{makecell}
\usepackage[table,xcdraw]{xcolor}
\usepackage{rotating}
\usepackage{threeparttable}
\usepackage{anyfontsize}
\usepackage{hyperref}
\usepackage[acronym]{glossaries}
\glsdisablehyper
\usepackage[capitalize,sort]{cleveref}
\crefname{figure}{Figure}{Figures}
\crefname{table}{Table}{Tables}

\usepackage[switch]{lineno}       %
\usepackage[
  style=nejm,
  citestyle=numeric-comp,
  sorting=none,
  maxcitenames=2,
  mincitenames=1,
  natbib=true
]{biblatex}
\newacronym{scrnaseq}{scRNA-seq}{single-cell RNA sequencing}
\newacronym{cdna}{cDNA}{complementary DNA}
\newacronym{de}{DE}{differential expression}
\newacronym{deg}{DEG}{differentially expressed gene}
\newacronym{dl}{DL}{deep learning}
\newacronym{dnn}{DNN}{deep neural network}
\newacronym{gan}{GAN}{generative adversarial network}
\newacronym{gnn}{GNN}{graph neural network}
\newacronym{ae}{AE}{autoencoder}
\newacronym{qc}{QC}{quality control}
\newacronym{geo}{GEO}{Gene Expression Omnibus}
\newacronym{hca}{HCA}{Human Cell Atlas}
\newacronym{em}{EM}{expectation-maximization}
\newacronym{zinb}{ZINB}{zero-inflated negative binomial}
\newacronym{pca}{PCA}{principal component analysis}
\newacronym{mst}{MST}{minimum spanning tree}
\newacronym{1d-cnn}{1D-CNN}{1D convolutional neural network}
\newacronym{encode}{ENCODE}{Encyclopedia of DNA Elements}

\newacronym{mae}{MAE}{mean absolute error}
\newacronym{medae}{MedAE}{median absolute error}
\newacronym{lnd}{LND}{log normalized difference}
\newacronym{mse}{MSE}{mean squared error}
\newacronym{pcc}{PCC}{pseudo-bulk correlation coefficient}
\newacronym{mcc}{MCC}{median correlation coefficient}
\newacronym{ari}{ARI}{adjusted rand index}
\newacronym{sc}{SC}{silhouette coefficient}
\newacronym{scc}{SCC}{Spearman's rank correlation coefficient}
\newacronym{nmi}{NMI}{normalized mutual information}
\newacronym{iou}{IoU}{intersection over union}
\newacronym{acc}{ACC}{macro accuracy}
\newacronym{pr}{PR}{macro precision}
\newacronym{rc}{RC}{macro recall}
\newacronym{f1}{F1}{macro F1 score}
\newacronym{pos}{POS}{pseudo-temporal ordering score}
\newacronym{krcc}{KRCC}{Kendall's rank correlation coefficient}
\newacronym{fpdeg}{FPDEG}{false positive DEG}

\newacronym{iqr}{IQR}{interquartile range}
\newacronym{umap}{UMAP}{uniform manifold approximation and projection}
\newacronym{t-sne}{t-SNE}{t-distributed stochastic neighbor embedding}
\newacronym{ddpm}{DDPM}{denoising diffusion probabilistic model}
\newacronym{knn}{$k$-NN}{$k$-nearest neighbor}
\newacronym{umi}{UMI}{unique molecular identifier}
\newacronym{lfc}{LFC}{log-fold change}

\title{One Method Does Not Fit All: A Large-Scale Benchmark of Imputation Methods for Single-Cell RNA Sequencing Data}

\author[1,2,3*$\dag$]{Yuichiro Iwashita}
\author[1,2$\dag$]{Ahtisham Fazeel Abbasi}
\author[2,3]{Koichi Kise}
\author[1,2,3,4]{\\Andreas Dengel}
\author[2,3,4]{Muhammad Nabeel Asim}

\affil[1]{RPTU University Kaiserslautern-Landau, Department of Computer Science, 67663 Kaiserslautern, Germany.}
\affil[2]{German Research Center for Artificial Intelligence (DFKI GmbH), 67663 Kaiserslautern, Germany.}
\affil[3]{Osaka Metropolitan University, Osaka 599-8531, Japan.}
\affil[4]{intelligentX GmbH (intelligentX.com), 67663 Kaiserslautern, Germany.}
\affil[*]{Address correspondence to: yuichiro.iwashita@cs.rptu.de}
\affil[$\dag$]{These authors contributed equally to this work.}

\date{}

\begin{document}
\twocolumn[
\begin{@twocolumnfalse}
\setstretch{1.0}

\maketitle

\begin{abstract}
Single-cell RNA sequencing (scRNA-seq) enables gene expression profiling at cellular resolution but is inherently affected by sparsity caused by dropout events, where expressed genes are recorded as zeros due to technical limitations. These dropout events distort gene expression distributions and compromise downstream analyses. To address this challenge, numerous traditional and deep learning (DL)-based imputation methods have been developed to detect dropout events and impute their true expression values. However, evaluations of these methods have largely been restricted to specific downstream tasks and limited datasets, and large-scale systematic comparisons under standardized benchmarking conditions are lacking. To fill this gap, our study presents a comprehensive framework that evaluates 16 imputation methods across 10 experimental protocols, 6 downstream tasks, and 30 benchmark datasets. Through extensive experimentation, we systematically investigate the strengths, limitations, and task-specific performance of these methods under consistent evaluation settings. Our results show that traditional methods perform relatively better across a wide range of downstream tasks, whereas the performance of DL-based methods is more variable. In addition, strong performance in numerical gene expression recovery does not necessarily translate into improved biological interpretability in downstream analyses, including cell clustering, differential expression analysis, marker gene analysis, trajectory analysis, and cell type annotation. Furthermore, method performance varies substantially across datasets, protocols, and downstream analyses, with no single method consistently outperforming the others. Our findings provide practical guidance for selecting imputation methods tailored to specific analytical objectives and underscore the importance of task-specific evaluation when assessing imputation performance in scRNA-seq data analysis.
\end{abstract}
\vspace{1em}
\end{@twocolumnfalse}
]

\glsresetall

\section{Introduction}\label{sec:introduction}

\Gls{scrnaseq} has become a powerful technology for profiling gene expression at the resolution of individual cells~\cite{tang2009mrnaseq,luecken2019current,rafi2025systematic,cheng2023review}. In \gls{scrnaseq}, individual cells are isolated from a tissue, and their mRNA content is reverse-transcribed into \gls{cdna}, amplified to increase signal, and sequenced to generate large collections of short DNA reads~\cite{hwang2018singlecell,kolodziejczyk2015technology,jovic2022singlecell,zheng2017massively,wen2026singlecell,hu2021nextgeneration}. These reads are subsequently processed through a computational pipeline that includes alignment to a reference genome, quality filtering, and transcript counting~\cite{hwang2018singlecell,kolodziejczyk2015technology,jovic2022singlecell,zheng2017massively}. The output of this pipeline is a structured gene expression matrix in which rows represent individual cells, columns represent genes or vice versa, and each matrix entry represents the expression of a specific gene in a cell~\cite{luecken2019current}.

\gls{scrnaseq} plays a significant role in biological research by addressing key questions related to cellular heterogeneity and disease mechanisms~\cite{hwang2018singlecell,luecken2019current}. For instance, \gls{scrnaseq} enables the discovery of dynamic gene regulatory features~\cite{kanton2019organoid}, the investigation of cellular interactions~\cite{ramilowski2015draft}, and the identification of rare cell types~\cite{huang2024advancement}. Moreover, \gls{scrnaseq} is an essential tool for constructing cell atlases, such as the \gls{hca}~\cite{regev2017human} and the Tabula Sapiens~\cite{thetabulasapiensconsortium2022tabula}, and these atlases enable comprehensive mapping of cell types and states across various tissues and organs~\cite{lahnemann2020eleven,huang2024advancement}.

A wide range of downstream tasks can be performed on \gls{scrnaseq} data to address diverse biological questions~\cite{luecken2019current,cheng2023evaluating,dai2022scimc,hou2020systematic,saelens2019comparison}. These tasks include cell clustering to group cells with similar expression profiles~\cite{traag2019louvain,blondel2008fast}, trajectory inference to model dynamic cellular processes~\cite{haghverdi2016diffusion,trapnell2014dynamics,bendall2014singlecell,street2018slingshot}, marker gene analysis to identify genes that define specific cell populations~\cite{pullin2024comparison}, cell type identification to assign biological identities to cells~\cite{wagner2016revealing}, and \gls{de} analysis to detect genes that are differentially expressed between different conditions~\cite{scholtens2005analysis}.

Despite the widespread use of \gls{scrnaseq}, the reliability of results from downstream tasks critically depends on the quality of gene expression data~\cite{jia2017accounting,andrews2018identifying}. In practice, achieving high-quality data is challenging due to substantial technical noise inherent to \gls{scrnaseq} experiments, arising from the limited amount of mRNA in individual cells~\cite{marinov2014singlecell}, inefficiencies in reverse transcription~\cite{islam2014quantitative}, and stochastic variability introduced during mRNA capture and amplification steps~\cite{islam2014quantitative,marinov2014singlecell,lahnemann2020eleven}. These technical limitations can lead to dropout events, where genes are observed as zero despite being expressed at low levels in the cell~\cite{lahnemann2020eleven,islam2014quantitative,marinov2014singlecell,kharchenko2014bayesian,wang2022imputation}. However, not all zero counts arise from technical noise; zero counts in \gls{scrnaseq} data may also reflect a true biological absence of transcription, often referred to as biological zeros, which are fundamentally distinct from technical dropout events~\cite{lahnemann2020eleven,jiang2022statistics,wang2022imputation}. Since dropout events distort the observed gene expression distribution, they can adversely affect the accuracy and robustness of downstream analyses~\cite{lahnemann2020eleven,wang2022imputation,stegle2015computational,chen2019singlecell}. In light of these limitations, data imputation strategies have been introduced to address dropout events before performing downstream analyses~\cite{luecken2019current,cheng2023evaluating,dai2022scimc,hou2020systematic,wang2022imputation}.

\gls{scrnaseq} imputation methods fall into traditional and \gls{dl}-based categories~\cite{zhang2025pbimpute,li2018accurate,huang2018saver,zhang2025acimpute,zhang2025sctsi,dijk2018recovering,zhang2021imputing,pan2021sclrtc,hu2021wedge,zhang2025scidpms,li2025scstd,zhang2025cpari,wang2023scmultigan,chen2023bubble,wang2021scgnn,xu2020scigans}. Traditional imputation methods typically rely on statistical modeling or similarity-based heuristics~\cite{zhang2025pbimpute,li2018accurate,huang2018saver,zhang2025acimpute,zhang2025sctsi,dijk2018recovering,zhang2021imputing,pan2021sclrtc,hu2021wedge}. Traditional methods can be broadly categorized into 3 methodological classes, namely model-based, smoothing-based, and low-rank matrix-based methods~\cite{lahnemann2020eleven,hou2020systematic,wang2022imputation}, and are briefly described in \cref{subsubsec:methods_traditional}. In contrast, \gls{dl}-based methods rely on representation learning using \glspl{dnn}~\cite{zhang2025scidpms,li2025scstd,zhang2025cpari,wang2023scmultigan,chen2023bubble,wang2021scgnn,xu2020scigans}, which is a fundamentally different approach compared to traditional methods. \gls{dl}-based methods can be broadly categorized into 4 methodological classes, namely diffusion-based, \gls{gan}-based, \gls{gnn}-based, and \gls{ae}-based methods, and are briefly discussed in \cref{subsubsec:methods_dl}.

\input{tables/existing_benchmarks}

Despite the availability of numerous imputation methods, existing benchmarking studies remain limited in their coverage of methods, datasets, experimental protocols, and downstream tasks. \cref{tab:related_work} summarizes 3 existing benchmarking studies on \gls{scrnaseq} data imputation, namely \citet{hou2020systematic}, \citet{dai2022scimc}, and \citet{cheng2023evaluating}. \citet{hou2020systematic} evaluated 12 traditional methods and 6 \gls{ae}-based methods, but do not include any diffusion-based, \gls{gan}-based, or \gls{gnn}-based methods. \citet{dai2022scimc} expanded the scope of \gls{dl}-based methods by incorporating 1 \gls{gan}-based and 1 \gls{gnn}-based methods in addition to 3 \gls{ae}-based methods. However, their evaluation is limited to 8 datasets and a single protocol. \citet{cheng2023evaluating} assessed 7 traditional methods and 4 \gls{ae}-based methods, but similarly, they do not evaluate diffusion-based, \gls{gan}-based, or \gls{gnn}-based methods. Furthermore, all 3 studies restricted their evaluation to at most 4 downstream tasks, which may not adequately capture the multifaceted effects of imputation on biological analyses. These gaps highlight the need for a more comprehensive and robust benchmarking study that spans a wider range of imputation methods, including recent \gls{dl} architectures, and evaluates their impact across a broader set of downstream tasks and protocols.

In this study, we address these limitations by presenting a comprehensive and systematic benchmark of \gls{scrnaseq} data imputation methods. Our evaluation covers 16 representative methods spanning 7 methodological categories, including both traditional methods (model-based, smoothing-based, and low-rank matrix-based) and recent \gls{dl}-based methods (diffusion-based, \gls{gan}-based, \gls{gnn}-based, and \gls{ae}-based methods). To ensure a robust and representative assessment, we evaluate these methods across 30 datasets (26 real and 4 simulated) sourced from 10 distinct protocols. Beyond imputing \gls{scrnaseq} data using these methods, we further investigate the impact of imputation on a broad range of biologically relevant downstream analyses. Specifically, we assess method performance across 6 key tasks in \gls{scrnaseq} data analysis, including numerical gene expression recovery, cell clustering, \gls{de} analysis, marker gene analysis, trajectory analysis, and cell type annotation. Together, this study provides a comprehensive benchmarking framework for evaluating \gls{scrnaseq} data imputation methods and offers practical insights into their strengths and limitations across diverse analytical settings. Our systematic comparison of methods across heterogeneous datasets, protocols, and downstream tasks provides guidance for selecting appropriate imputation strategies tailored to specific single-cell analysis objectives.

\section{Materials and Methods}

\subsection{Experimental Design}

\begin{figure*}[p]
    \centering
    \includegraphics[width=\textwidth,height=0.92\textheight,keepaspectratio]{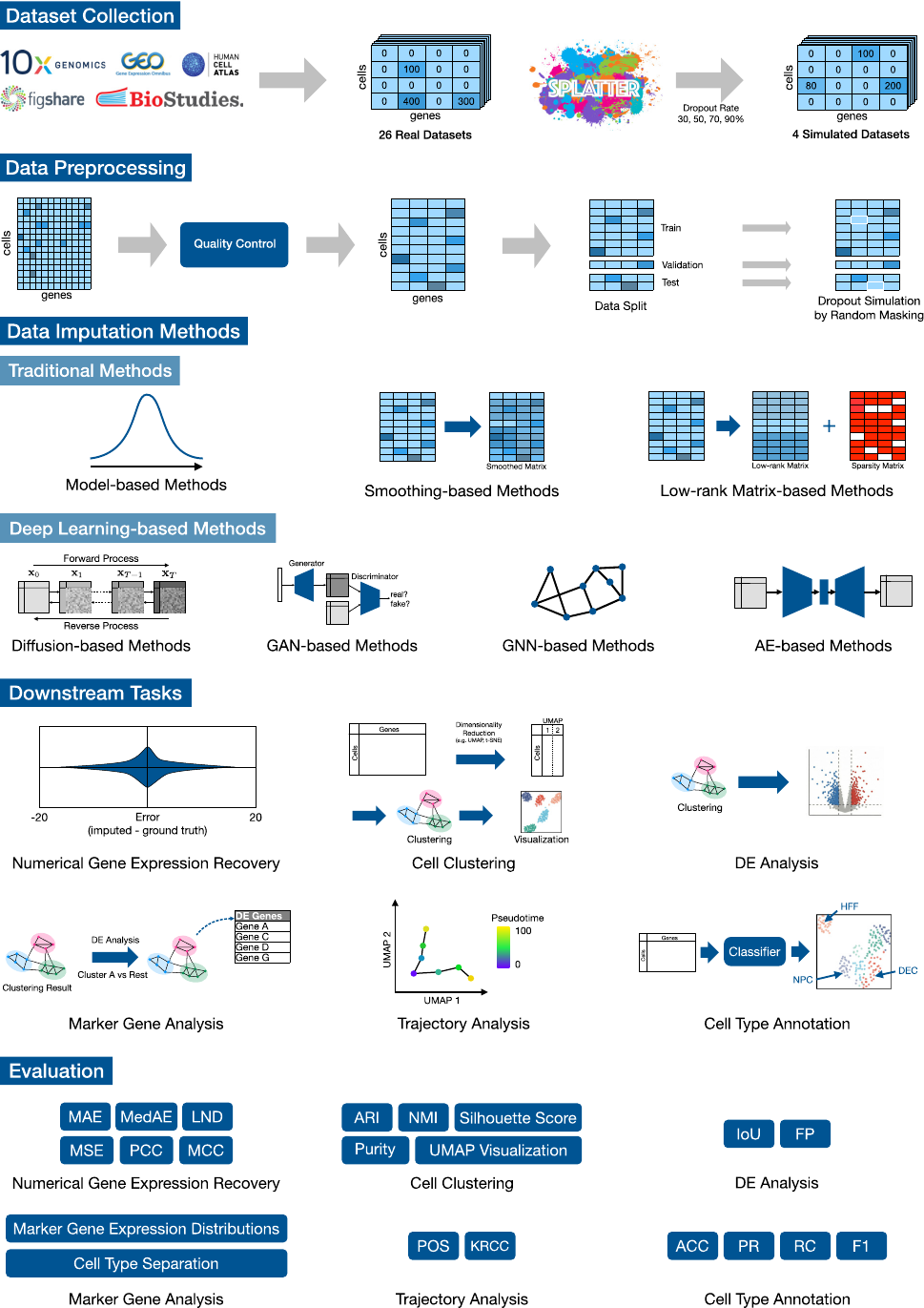}
    \caption{Data Imputation Benchmarking Framework.}
    \label{fig:framework}
\end{figure*}

\gls{scrnaseq} data cannot be reliably analyzed directly due to the presence of excessive dropout events, which manifest as false zero expression values and distort observed gene expression distributions~\cite{luecken2019current}. To address this limitation, data imputation methods aim to recover latent gene expression signals and improve the robustness of downstream tasks~\cite{cheng2023evaluating,dai2022scimc,hou2020systematic,wang2022imputation,li2018accurate,zhang2025pbimpute,dijk2018recovering,zhang2025sctsi,zhang2025acimpute,zhang2021imputing,pan2021sclrtc,hu2021wedge,huang2018saver,zhang2025scidpms,li2025scstd,wang2023scmultigan,xu2020scigans,wang2021scgnn,chen2023bubble,zhang2025cpari}. In this study, a comprehensive benchmarking framework is designed to systematically evaluate a diverse set of \gls{scrnaseq} imputation methods, spanning 9 traditional methods, including model~\cite{li2018accurate,zhang2025pbimpute,huang2018saver}, smoothing~\cite{dijk2018recovering,zhang2025sctsi,zhang2025acimpute}, and low-rank matrix-based methods~\cite{zhang2021imputing,pan2021sclrtc,hu2021wedge}, and 7 \gls{dl}-based methods, including diffusion~\cite{zhang2025scidpms,li2025scstd}, \gls{gan}~\cite{wang2023scmultigan,xu2020scigans}, \gls{gnn}~\cite{wang2021scgnn}, and \gls{ae}-based methods~\cite{chen2023bubble,zhang2025cpari}. These methods differ substantially in their underlying assumptions, including statistical modeling of dropout mechanisms~\cite{li2018accurate,zhang2025pbimpute}, exploiting cell-cell similarity~\cite{dijk2018recovering,zhang2025sctsi,zhang2025acimpute}, enforcing low-rank structures~\cite{zhang2021imputing,pan2021sclrtc,hu2021wedge}, or learning latent representations through \glspl{dnn}~\cite{zhang2025scidpms,zhang2025cpari,li2025scstd,wang2023scmultigan,chen2023bubble,wang2021scgnn,xu2020scigans}. Given this diversity, it is essential to assess their effectiveness across heterogeneous datasets and multiple downstream tasks rather than relying on a single evaluation criterion~\cite{cheng2023evaluating,dai2022scimc,hou2020systematic}.

To ensure a fair and unbiased comparison, the benchmarking framework incorporates a carefully curated collection of 26 real and 4 simulated \gls{scrnaseq} datasets that vary in size, sparsity level, biological context, and protocol. Each dataset undergoes standardized \gls{qc} and is split into training, validation, and test sets to prevent data leakage during training and evaluation. Dropout events are artificially introduced into each set to establish known ground truth values, enabling objective assessment of numerical gene expression recovery~\cite{cheng2023evaluating,dai2022scimc,hou2020systematic}. This design allows the framework to isolate the true impact of each imputation method on data quality while avoiding overfitting and biased performance estimates.

Evaluation of data imputation methods is performed from both numerical and functional perspectives using 6 downstream tasks. First, the performance of numerical gene expression recovery is quantified by directly comparing imputed and ground truth values using error-based metrics~\cite{cheng2023evaluating,dai2022scimc}. Subsequently, biological utility is assessed through a diverse set of downstream tasks, including cell clustering~\cite{traag2019louvain,blondel2008fast,kiselev2019challenges}, \gls{de} analysis~\cite{luecken2019current,scholtens2005analysis}, marker gene analysis~\cite{luecken2019current,pullin2024comparison}, trajectory analysis~\cite{haghverdi2016diffusion,trapnell2014dynamics,bendall2014singlecell,street2018slingshot,luecken2019current}, and cell type annotation~\cite{wagner2016revealing,luecken2019current,pasquini2021automated,kiselev2019challenges}. These tasks collectively capture core analytical objectives in \gls{scrnaseq} studies and provide insight into how data imputation influences biological interpretation~\cite{cheng2023evaluating,dai2022scimc,hou2020systematic}. Importantly, all downstream tasks are conducted consistently across methods to ensure comparability.

\cref{fig:framework} presents an overview of the proposed benchmarking framework, illustrating the complete pipeline from dataset collection and preprocessing to data imputation, downstream tasks, and performance evaluation.
Detailed descriptions of dataset collection and preprocessing are provided in \cref{subsec:method_datasets}.
Data imputation methods, downstream tasks, and evaluation measures are described in \cref{subsec:methods_imputation,subsec:method_downstream_tasks,subsec:method_evaluation_metrics}, respectively.
Together, this framework enables a systematic and reproducible assessment of \gls{scrnaseq} data imputation methods, highlighting their strengths, limitations, and suitability for different analytical scenarios.

\subsection{Benchmark Datasets}
\label{subsec:method_datasets}

The performance of \gls{scrnaseq} imputation methods is heavily influenced by the characteristics of the datasets, such as dataset size, sparsity rate, biological context, and protocol~\cite{cheng2023evaluating,dai2022scimc,hou2020systematic}.
In particular, datasets not only affect the difference between imputed and ground truth expression values but also directly impact downstream tasks~\cite{cheng2023evaluating,dai2022scimc,hou2020systematic}, including cell clustering~\cite{traag2019louvain,blondel2008fast,kiselev2019challenges}, \gls{de} analysis~\cite{luecken2019current,scholtens2005analysis}, marker gene analysis~\cite{luecken2019current,pullin2024comparison}, trajectory analysis~\cite{haghverdi2016diffusion,trapnell2014dynamics,bendall2014singlecell,street2018slingshot,luecken2019current}, and cell type annotation~\cite{wagner2016revealing,luecken2019current,pasquini2021automated,kiselev2019challenges}.
Therefore, the use of representative, well-curated, and high-quality \gls{scrnaseq} datasets is fundamental to conducting an unbiased and meaningful benchmark~\cite{cheng2023evaluating,hou2020systematic}.

\input{tables/datasets}

\cref{tab:dataset_details} presents 30 unique benchmark datasets spanning 10 \gls{scrnaseq} data extraction protocols, 13 cell lines, 10 tissues from human and mouse samples, and 6 disease conditions which are collected from \href{https://www.10xgenomics.com/datasets}{\mbox{10x Genomics}} \href{https://www.10xgenomics.com/datasets}{\mbox{dataset repository}}~\cite{10x}, \href{https://www.ncbi.nlm.nih.gov/geo}{\gls{geo}} database~\cite{edgar2002gene}, \href{https://figshare.com}{Figshare}~\cite{figshare}, \href{https://data.humancellatlas.org/}{\gls{hca}}~\cite{regev2017human} and \href{https://www.ebi.ac.uk/biostudies}{BioStudies}~\cite{sarkans2018biostudies}. These datasets are selected for representative coverage of the heterogeneity and protocol variations present in real-world \gls{scrnaseq} studies.

The curated dataset collection covers a wide range of sizes and sparsity rates to evaluate imputation methods under different number of cells, genes, and dropout conditions. These datasets vary significantly in scale, with number of cells ranging from 224 to 18,177 and number of genes spanning 10,850 to 36,092. This variety helps evaluate if data imputation methods can handle both small samples and large-scale data~\cite{cheng2023evaluating}. Furthermore, the collection includes sparsity rates from $45.02\,\%$ to $95.75\,\%$, reflecting realistic dropout levels encountered in \gls{scrnaseq} experiments~\cite{cheng2023evaluating}. Conducting benchmarks across these different levels of data density ensures a clear evaluation of how well each method recovers gene expression values in both high and low-quality datasets~\cite{cheng2023evaluating,hou2020systematic}.

Protocols introduce distinct technical noise profiles and dropout patterns that can substantially influence imputation performance~\cite{kolodziejczyk2015technology,jovic2022singlecell,zheng2017massively}. This protocol-level heterogeneity is addressed by selecting 26 datasets obtained using 10 different protocols, namely 10x Chromium~\cite{zheng2017massively}, SMART-seq~\cite{ramskold2012fulllength}, SMART-seq2~\cite{picelli2014fulllength}, CEL-seq2~\cite{hashimshony2016celseq2}, Drop-seq~\cite{macosko2015highly}, inDrop~\cite{klein2015droplet}, Microwell-seq~\cite{han2018mapping}, STRT-seq~\cite{islam2012highly}, Sort-seq~\cite{muraro2016singlecell}, and Fluidigm C1~\cite{xin2016use}. This dataset collection enables a systematic assessment of imputation robustness across platforms with fundamentally different library preparation strategies, read depths, and technical noise profiles~\cite{cheng2023evaluating,hou2020systematic}.

In addition to real datasets, 4 simulated datasets with known ground truth to compare the performance of imputation methods under different dropout rates are used in this benchmarking framework~\cite{cheng2023evaluating}. The simulated datasets are generated using the \href{https://github.com/Oshlack/splatter}{Splatter}~\cite{zappia2017splatter} package, which allows controlled simulation of \gls{scrnaseq} data with known ground truth~\cite{cheng2023evaluating,zappia2017splatter}. Each simulated dataset contains 2,000 cells, 600 genes, and 5 clusters. To compare the performance of imputation methods under different dropout rates, 4 simulated datasets with varying dropout rates of $30.79$, $50.62$, $70.12$, and $89.61\,\%$ are generated, which reflect a range of dropout conditions commonly observed in real \gls{scrnaseq} experiments~\cite{cheng2023evaluating}.

Furthermore, following \citet{hou2020systematic}, corresponding bulk RNA-seq datasets are obtained for 3 real \gls{scrnaseq} datasets, namely sc\_10x\_5cl, encode\_fluidigm\_5cl, and hca\_10x\_tissue, to evaluate imputation performance at the cell population level. These bulk RNA-seq datasets are downloaded from the processed data repository\footnote{\url{https://github.com/Winnie09/imputationBenchmark}} provided by \citet{hou2020systematic}. For sc\_10x\_5cl, bulk RNA-seq data with 10 samples of 5 cell lines, namely HCC827, H1975, H2228, H838, and A549, are originally collected from GSE86337~\cite{holik2017rna,hou2020systematic}. For encode\_fluidigm\_5cl, bulk RNA-seq data with 58 samples of 5 cell lines, namely A549, GM12878, H1-hESC, IMR90, and K562, are originally collected from the \gls{encode}~\cite{2004encode,hou2020systematic}. For hca\_10x\_tissue, bulk RNA-seq data with 49 samples of 13 cell types, namely B cell, CD4 T cell, CD8 T cell, CMP, GMP, HSC, MEP, monocyte, MPP, NK cells, CLP, erythroid, and LMPP, are originally collected from GSE74246~\cite{corces2016lineage,hou2020systematic}.

After dataset collection, \gls{qc} is applied to each real dataset to remove low-quality cells and genes~\cite{cheng2023evaluating}. Following \citet{cheng2023evaluating}, cells whose number of expressed genes or total count is larger than the 75th percentile plus 3 times the \gls{iqr} or less than the 25th percentile minus 3 times the \gls{iqr} across all cells are filtered out~\cite{cheng2023evaluating}. Similarly, genes which are expressed in fewer than 3 cells are filtered out~\cite{cheng2023evaluating}. With this \gls{qc} procedure, only high-quality cells and genes are retained for subsequent imputation and downstream tasks~\cite{cheng2023evaluating}.

The cells are randomly partitioned into training, validation, and test sets with a ratio of $70\,\%$, $10\,\%$, and $20\,\%$, respectively. The methods that train a model before imputation are trained on the training and validation sets and then impute all cells, where scIDPMs$^\dagger$ and scSTD use the validation set for early stopping, and the other methods process all cells at once as in their original implementations (Supplementary Table S7). The test set is used for evaluating the imputation methods, including running downstream tasks. For the methods trained before imputation, this splitting strategy keeps the test cells out of training, whereas previous benchmarking studies~\cite{hou2020systematic,dai2022scimc,cheng2023evaluating} use the same ground truth data for both training and testing, which can lead to overfitting and biased results.

After \gls{qc} and data splitting, dropout events are artificially introduced into each split set of real \gls{scrnaseq} datasets since it is not possible to distinguish between true biological zeros and technical dropout events in real \gls{scrnaseq} datasets~\cite{lahnemann2020eleven,jiang2022statistics,wang2022imputation}. To introduce dropout events, following \citet{cheng2023evaluating}, $10\,\%$ of non-zero expression values in each split dataset are selected and masked as zero expression values~\cite{cheng2023evaluating}. The original non-zero expression values before masking are used as ground truth for evaluation.

\subsection{Data Imputation Methods}
\label{subsec:methods_imputation}

Imputation methods are included in this benchmark when they meet all of the following 3 criteria. First, the method is designed for imputing dropout events in \gls{scrnaseq} data. Second, the method is published in a peer-reviewed journal between 2020 and 2025, except for scImpute~\cite{li2018accurate}, SAVER~\cite{huang2018saver}, and MAGIC~\cite{dijk2018recovering}, which are published in 2018 and included as widely used reference methods. Third, an implementation of the method is publicly available. As a result, 16 methods are selected, namely 9 traditional methods in 3 categories and 7 \gls{dl}-based methods in 4 categories, as described below.

\subsubsection{Traditional Methods}
\label{subsubsec:methods_traditional}

Traditional data imputation methods for \gls{scrnaseq} can be broadly categorized into 3 methodological classes: model-based methods, smoothing-based methods, and low-rank matrix-based methods. These categories and the specific methods evaluated in this study are described below.

\paragraph{Model-based Methods.} Model-based methods explicitly model the occurrence of dropout events and the distribution of gene expression values using parametric statistical models to estimate and impute dropout events in \gls{scrnaseq} data~\cite{li2018accurate,zhang2025pbimpute,huang2018saver}. 3 different model-based methods are selected in this study, i.e., scImpute~\cite{li2018accurate}, PbImpute~\cite{zhang2025pbimpute}, and SAVER~\cite{huang2018saver}. scImpute~\cite{li2018accurate} models the occurrence of dropout events and the distribution of gene expression values using a mixture distribution, in which dropout events are modeled by a Gamma distribution and true expression values are modeled by a normal distribution~\cite{li2018accurate}. An \gls{em} algorithm is then used to estimate the dropout probability of each zero expression value~\cite{dempster1977maximum,li2018accurate}. Cells with similar expression patterns are then identified using non-negative least squares regression, and expression values from these similar cells are used to impute values at inferred dropout positions~\cite{li2018accurate}. In contrast, PbImpute~\cite{zhang2025pbimpute} addresses the common problem of over-imputation by utilizing a multi-stage method~\cite{zhang2025pbimpute}. The process begins with \gls{zinb} modeling to provide robust dropout identification and initial imputation~\cite{zhang2025pbimpute}. To enhance data fidelity, the method incorporates a static repair step that corrects over-imputed values by adjusting outlying nonzero values~\cite{zhang2025pbimpute}. Moreover, the method identifies residual dropout events using node2vec~\cite{grover2016node2vec}, which capture complex relationships between cells, and impute residual dropout events dynamically~\cite{zhang2025pbimpute}. This multi-stage approach allows PbImpute to accurately identify and impute dropout events while minimizing the risk of over-imputation~\cite{zhang2025pbimpute}. SAVER~\cite{huang2018saver} recovers the true expression level of each gene in each cell by borrowing information across genes~\cite{huang2018saver}. SAVER models the observed count of each gene in each cell by a Poisson-Gamma mixture, i.e., a negative binomial distribution, whose Gamma prior represents the true expression level~\cite{huang2018saver}. The prior mean is estimated in an empirical Bayes-like manner by a penalized Poisson LASSO regression that predicts the expression of the gene from the expression of other genes, and the prior variance is estimated by assuming a gene-specific dispersion that is constant across cells~\cite{huang2018saver}. SAVER then outputs the posterior distribution of the true expression, and the posterior mean is used as the imputed value~\cite{huang2018saver}.

\paragraph{Smoothing-based Methods.} Smoothing-based methods leverage the similarity of expression profiles among neighboring cells to reduce noise and preserve biological signals in \gls{scrnaseq} data~\cite{dijk2018recovering,zhang2025sctsi,zhang2025acimpute}, a process typically referred to as smoothing~\cite{dijk2018recovering,zhang2025sctsi,zhang2025acimpute}. 3 representative smoothing-based methods are selected in this study, i.e., MAGIC~\cite{dijk2018recovering}, scTsI~\cite{zhang2025sctsi}, and AcImpute~\cite{zhang2025acimpute}. MAGIC is designed based on the concept that gene expression profiles can be shared among similar cells to recover missing values, and utilizes Markov processes to impute dropout events~\cite{dijk2018recovering}. MAGIC first calculates the cell-cell Euclidean distance matrix and constructs a cell-cell affinity matrix using a Gaussian kernel~\cite{dijk2018recovering}. The affinity matrix is then normalized using row normalization to obtain a Markov transition matrix, which represents the transition probabilities between cells~\cite{dijk2018recovering}. The process of calculating the Markov matrix is repeated multiple times to reduce noise and keep the biological signal~\cite{dijk2018recovering}. Finally, the imputed gene expression matrix is obtained by multiplying the final Markov transition matrix with the observed gene expression matrix~\cite{dijk2018recovering}. scTsI is a 2-stage smoothing-based imputation method that first imputes the zero expression values using the information of neighboring cells and genes, and then adjusts the imputed values using ridge regression~\cite{mcdonald2009ridge,zhang2025sctsi}. In the first stage, scTsI imputes the zero expression values by calculating an average of the expression values from nearest neighbor cells and nearest neighbor genes~\cite{zhang2025sctsi}. In the second stage, scTsI refines the imputed values by fitting a ridge regression model that predicts the expression value of each gene in each cell based on the expression values of other genes in the same cell and the same gene in other cells~\cite{zhang2025sctsi}. While MAGIC diffuses gene expression values equally across all genes, AcImpute applies different diffusion strengths for highly and lowly expressed genes based on the observation that dropout events are more prevalent in lowly expressed genes~\cite{zhang2025acimpute}. AcImpute first normalizes, selects highly variable genes, and reduces dimensionality using \gls{pca}~\cite{pearson1901liii,zhang2025acimpute}. Similar to MAGIC, AcImpute then constructs a cell-cell affinity matrix using \gls{knn}-based adaptive kernel, and obtains a Markov transition matrix through row normalization~\cite{zhang2025acimpute}. In addition, AcImpute calculates the power matrix, a locally averaged diffusion operator, by averaging the normalized matrix over its neighboring cells to capture average gene expression patterns across neighboring cells~\cite{zhang2025acimpute}. Finally, AcImpute combines the Markov transition matrix, the power matrix, and the observed gene expression matrix to create a modified Markov transition matrix~\cite{zhang2025acimpute}.

\paragraph{Low-rank Matrix-based Methods.}
Low-rank matrix-based methods are built on the idea that gene expression data often have an underlying simple structure~\cite{zhang2021imputing,pan2021sclrtc,hu2021wedge}. Specifically, these methods assume that the true gene expression matrix can be well-approximated by a matrix with low rank, meaning that the expression patterns of thousands of genes across many cells can actually be explained by a small number of shared biological factors, such as cell types or cell states~\cite{zhang2021imputing,pan2021sclrtc,hu2021wedge}. In practice, \gls{scrnaseq} data contain a large number of zero values, some of which are caused by dropout events rather than true absence of gene expression~\cite{lahnemann2020eleven}. To address this, the observed gene expression matrix can be formulated as $X_{\mathrm{obs}} = X_{\mathrm{true}} + E \in \mathbb{R}^{n \times m}$, where $n$ is the number of cells, $m$ is the number of genes, $X_{\mathrm{true}}$ is the true gene expression matrix to be estimated, and $E$ is a sparse noise matrix that captures dropout events~\cite{zhang2021imputing,pan2021sclrtc,hu2021wedge}. The objective of low-rank matrix-based methods is to estimate $X_{\mathrm{true}}$ by using the low-rank structure of the gene expression data while accounting for dropout events represented by $E$~\cite{zhang2021imputing,pan2021sclrtc,hu2021wedge}. 3 representative low-rank matrix-based methods are selected in this study, i.e., PBLR~\cite{zhang2021imputing}, scLRTC~\cite{pan2021sclrtc}, and WEDGE~\cite{hu2021wedge}. PBLR explicitly incorporates cell heterogeneity into the imputation process~\cite{zhang2021imputing}. It first identifies cell subpopulations by constructing multiple cell-cell affinity matrices and applying non-negative matrix factorization followed by hierarchical clustering~\cite{zhang2021imputing}. This step partitions the global expression matrix into more homogeneous submatrices~\cite{zhang2021imputing}. For each subpopulation-specific sub-matrix, PBLR performs bounded low-rank matrix recovery, where dropout values are constrained by gene-specific upper bounds estimated from observed expression levels~\cite{zhang2021imputing}. This bounded formulation prevents unrealistically large imputations and improves recovery accuracy, especially in heterogeneous datasets~\cite{zhang2021imputing}. scLRTC generalizes matrix-based approaches by modeling \gls{scrnaseq} data as a third-order tensor, constructed using cell-cell similarity information~\cite{pan2021sclrtc}. This tensor representation enables simultaneous modeling of gene-gene and cell-cell correlations. scLRTC applies low-rank tensor completion to recover missing values, effectively denoising the data while preserving higher-order structural relationships~\cite{pan2021sclrtc}. By leveraging tensor decomposition rather than simple matrix factorization, scLRTC can better capture complex dependencies in \gls{scrnaseq} data and improve downstream analyses such as clustering and trajectory inference~\cite{pan2021sclrtc}. WEDGE addresses dropout by introducing a biased low-rank matrix decomposition framework~\cite{hu2021wedge}. Unlike standard matrix factorization methods that ignore zero entries, WEDGE assigns different weights to zero and non-zero elements in the objective function~\cite{hu2021wedge}. Non-zero entries are fitted closely to preserve observed expression, while zero entries are softly penalized using a tunable bias parameter, reducing the risk of over-imputation~\cite{hu2021wedge}. The model is optimized via alternating non-negative least squares, ensuring biologically meaningful imputed values~\cite{hu2021wedge}. This weighted strategy allows WEDGE to robustly recover expression patterns in highly sparse \gls{scrnaseq} datasets~\cite{hu2021wedge}.

\subsubsection{\texorpdfstring{\Gls{dl}-based Methods}{DL-based Methods}}
\label{subsubsec:methods_dl}

\gls{dl}-based methods can be broadly categorized into 4 methodological classes: diffusion, \gls{gan}, \gls{gnn}, and \gls{ae}-based methods. These categories and the specific methods evaluated in this study are described below.

\paragraph{Diffusion-based Methods.} Diffusion-based methods utilize diffusion models~\cite{sohl-dickstein2015deep,ho2020denoising} to model the underlying data distribution and impute dropout events in \gls{scrnaseq} data~\cite{zhang2025scidpms,li2025scstd}. Most diffusion models are built on \glspl{ddpm}~\cite{ho2020denoising} composed of 2 Markov processes, the forward process that gradually adds Gaussian noise to the data over multiple time steps, and the reverse process that learns to recover the original data from the noisy input step by step~\cite{ho2020denoising}. Moreover, conditional diffusion models can align the output of the reverse denoising process with the given conditions~\cite{ho2020denoising}. In this study, 2 representative diffusion-based methods are selected, i.e., scIDPMs~\cite{zhang2025scidpms} and scSTD~\cite{li2025scstd}. scIDPMs identifies potential dropout sites by leveraging intercellular relationships and trains a conditional \gls{ddpm} conditioned on the observed gene expression values~\cite{zhang2025scidpms}. To train scIDPMs, the method receives the imputation target matrix as input, where it represents the true values and the positions of dropout events, and the observed gene expression matrix as a condition, where it shows gene expression values of the remaining part, and learns the parameters by adding noise to the imputation target matrix and removing the noise from it~\cite{zhang2025scidpms}. During the inference step of scIDPMs, the method receives the imputation target matrix with random noise as input and the observed gene expression matrix as a condition, and outputs an estimated gene expression matrix corresponding to the imputation target matrix~\cite{zhang2025scidpms}. In contrast, scSTD~\cite{li2025scstd} combines an \gls{ae} with a latent diffusion model whose denoising network is based on the Swin Transformer~\cite{li2025scstd}. scSTD first trains an \gls{ae} on the gene expression matrix, and the encoder maps each cell to a compact latent embedding~\cite{li2025scstd}. A diffusion model is then trained on these latent embeddings to learn the distribution of cell states~\cite{li2025scstd}. During the inference step of scSTD, the method synthesizes realistic latent neighbors for each cell with the trained diffusion model and aggregates their decoded expression profiles to obtain the imputed gene expression of the cell~\cite{li2025scstd}.

\paragraph{\texorpdfstring{\Gls{gan}-based Methods}{GAN-based Methods}.} \Gls{gan}-based methods utilize \glspl{gan} to learn the underlying distribution of \gls{scrnaseq} data and generate imputed values~\cite{wang2023scmultigan,xu2020scigans}. \glspl{gan} are composed of 2 neural networks, a generator and a discriminator, that are trained in an adversarial manner~\cite{goodfellow2014generative}. The generator learns to generate realistic data samples from random noise, while the discriminator learns to distinguish between real and generated samples~\cite{goodfellow2014generative}. Through the training process, \glspl{gan} can learn complex data distributions and generate high-quality samples~\cite{goodfellow2014generative}. Due to their ability to model complex data distributions, \gls{gan}-based imputation methods are being proposed~\cite{wang2023scmultigan,xu2020scigans}. In this study, 2 representative \gls{gan}-based methods are included, i.e., scMultiGAN~\cite{wang2023scmultigan} and scIGANs~\cite{xu2020scigans}. scIGANs is designed to apply image-generating \glspl{gan} to \gls{scrnaseq} data~\cite{xu2020scigans}. scIGANs first converts \gls{scrnaseq} data to grayscale square images, which are the format accepted as input by image-generating \glspl{gan}, by reshaping gene expression vector of a cell into a grayscale square image~\cite{xu2020scigans}. The squared images are fed into a \gls{gan} and the model learns parameters by generating fake samples and distinguishing between the true samples and the fake samples~\cite{xu2020scigans}. During the inference step of scIGANs, the method generates synthetic grayscale square images from the observed \gls{scrnaseq} data, selects \gls{knn} cells of the cell to be imputed, and imputes based on the generated image~\cite{xu2020scigans}. While scIGANs simply uses a single \gls{gan} to generate synthetic cells, scMultiGAN utilizes 3 \glspl{gan} to learn the complex patterns of \gls{scrnaseq} data and generate high-quality imputed values~\cite{wang2023scmultigan}. scMultiGAN performs \gls{scrnaseq} imputation using multiple \glspl{gan} with a two-stage training strategy~\cite{wang2023scmultigan}. In the first stage of scMultiGAN, 2 \glspl{gan} are trained to learn the distribution of true expression values and dropout events separately~\cite{wang2023scmultigan}. To precisely impute dropout events, in the second stage, it learns the distribution of true expression values precisely by integrating the true expression values generator trained in the first stage, an additional U-Net~\cite{ronneberger2015unet}-based generator, and a discriminator~\cite{wang2023scmultigan}. Finally, the generator from the second stage is used to impute dropout events~\cite{wang2023scmultigan}.

\paragraph{\texorpdfstring{\Gls{gnn}-based Methods}{GNN-based Methods}.} \Gls{gnn}-based methods leverage \glspl{gnn} to model the relationships between cells and impute dropout events~\cite{wang2021scgnn}. By propagating information through the graph structure, \glspl{gnn} can aggregate neighborhood-level features and effectively capture both local and global cellular relationships~\cite{scarselli2009graph}. In this study, scGNN~\cite{wang2021scgnn} is included as a representative \gls{gnn}-based method. scGNN is a hypothesis-free \gls{gnn}-based method and it integrates 3 iterative multi-modal \glspl{ae}, namely feature \gls{ae}, graph \gls{ae}, cluster \gls{ae}, to model heterogeneous gene expression patterns and aggregate cell-cell relationships~\cite{wang2021scgnn}. The feature \gls{ae} receives the regularized gene expression matrix calculated through the left-truncated mixture Gaussian model~\cite{wan2019ltmg} as input and learns low-dimensional cell representations by minimizing the reconstruction loss between the input and output of the \gls{ae}~\cite{wang2021scgnn}. Based on the output of the feature \gls{ae}, scGNN constructs a cell-cell graph using \gls{knn} and feeds it into the graph \gls{ae} to aggregate neighborhood-level features and learn enhanced cell representations~\cite{wang2021scgnn}. The cluster \gls{ae} receives the reconstructed gene expression matrix from the feature \gls{ae} and an individual encoder is used for each cell cluster to better capture cluster-specific gene expression patterns, which are identified through clustering on the output of the graph \gls{ae}~\cite{wang2021scgnn}. The reconstructed gene expression matrices from an individual encoder of the cluster \gls{ae} are concatenated, and fed into the feature \gls{ae} and graph \gls{ae} in the next iteration~\cite{wang2021scgnn}. This iterative process continues until convergence, and the final reconstructed gene expression matrix from the feature \gls{ae} is used as the imputed gene expression matrix~\cite{wang2021scgnn}.

\paragraph{\texorpdfstring{\Gls{ae}-based Methods}{AE-based Methods}.} \Gls{ae}-based methods utilize \gls{ae} architectures to learn low-dimensional representations of \gls{scrnaseq} data and reconstruct imputed values~\cite{zhang2025cpari,chen2023bubble}. \Glspl{ae} are encoder-decoder architectures that consist of an encoder that maps the input data to a low-dimensional latent representation, and a decoder that reconstructs the original data from the latent representation~\cite{hinton2006reducing,kingma2022autoencoding}. By training the \gls{ae} to minimize the reconstruction loss, which measures the error between the ground truth and reconstructed data, \glspl{ae} can learn meaningful representations of the input data~\cite{hinton2006reducing,kingma2022autoencoding}. \Glspl{ae} are adapted for \gls{scrnaseq} data imputation due to their ability to capture complex gene expression patterns and reconstruct dropout events~\cite{zhang2025cpari,chen2023bubble}. In this study, 2 representative \gls{ae}-based methods are included, i.e., Bubble~\cite{chen2023bubble} and CPARI~\cite{zhang2025cpari}. Bubble utilizes an \gls{ae} to selectively impute dropout events that are identified through statistical analysis of gene expression patterns within cell subpopulations~\cite{chen2023bubble}. Bubble consists of 2 main steps, namely identification of dropout events, and imputation~\cite{chen2023bubble}. In the first step, Bubble first reduces the dimensionality of the observed gene expression matrix using \gls{pca}~\cite{pearson1901liii}, divides cells into clusters using $k$-means clustering, and identifies dropout events through predefined statistical rules, which state that if a gene has a high expression rate and low variation in cells within a cluster, then zero expression levels of the gene in the cluster are more likely to be dropout events~\cite{chen2023bubble}. In the second step, Bubble trains an \gls{ae} with the objective of minimizing the total loss function composed of reconstruction loss of the \gls{ae}, biological loss, which aims to recover non-zero expression values, and alignment loss, which aims to align the aggregated reconstructed gene expression values to the matched bulk RNA-seq data, to impute the identified dropout events~\cite{chen2023bubble}. On the other hand, CPARI combines cell partitioning with absolute and relative imputation strategies to effectively distinguish biological zeros from dropout events~\cite{zhang2025cpari}. In the first step, CPARI selects highly variable genes, and partitions cells into multiple clusters using fuzzy C-means clustering~\cite{dunn1973fuzzy,zhang2025cpari}. Absolute imputation is done for each cell cluster by identifying dropout events from the observed gene expression values by statistical rules, and imputing dropout events using an \gls{ae}~\cite{zhang2025cpari}. As absolute imputation alone may not fully identify and impute all dropout events, relative imputation is performed by statistical rules based on gene expression patterns within cell subpopulations~\cite{zhang2025cpari}. Finally, the outputs of absolute and relative imputation are integrated to create the final imputed gene expression matrix~\cite{zhang2025cpari}.

\subsection{Downstream Tasks}
\label{subsec:method_downstream_tasks}

\begin{figure*}[tb]
    \centering
    \includegraphics[width=\textwidth,keepaspectratio]{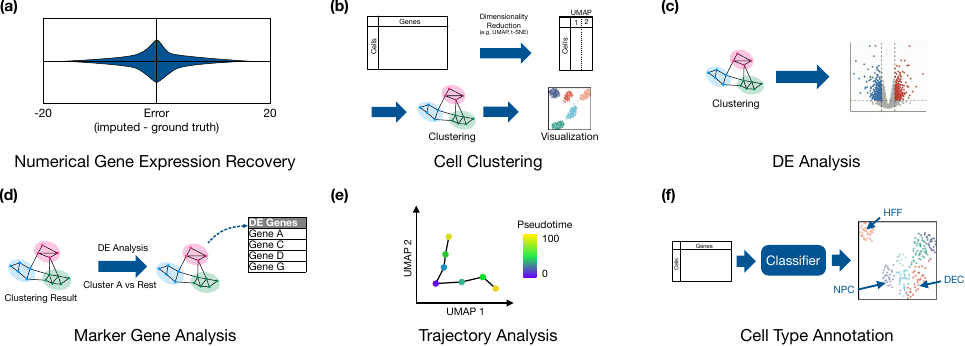}
    \caption{The overview of downstream tasks used for benchmarking imputation methods. \textbf{(a)} Numerical gene expression recovery, \textbf{(b)} cell clustering, \textbf{(c)} DE analysis, \textbf{(d)} marker gene analysis, \textbf{(e)} trajectory analysis, and \textbf{(f)} cell type annotation.}
    \label{fig:downstream_tasks}
\end{figure*}

The quality of imputed \gls{scrnaseq} data directly influences the reliability and performance of computational models in downstream tasks~\cite{cheng2023evaluating,hou2020systematic,dai2022scimc}. Inaccurate or biased imputation can distort underlying biological signals, leading to misleading conclusions~\cite{cheng2023evaluating,hou2020systematic,dai2022scimc}. Therefore, a comprehensive evaluation of imputation methods must assess not only numerical recovery of gene expression values but also their impact on biologically meaningful downstream tasks~\cite{cheng2023evaluating,hou2020systematic,dai2022scimc}. In this section, as illustrated in \cref{fig:downstream_tasks}, we describe 6 distinct downstream tasks used to benchmark imputation methods, namely numerical gene expression recovery, cell clustering, \gls{de} analysis, marker gene analysis, trajectory analysis, and cell type annotation.

\paragraph{Numerical Gene Expression Recovery.} Numerical gene expression recovery can be formulated as a regression task, in which the objective is to predict true gene expression values from sparsely observed \gls{scrnaseq} data affected by dropout events~\cite{cheng2023evaluating}. In this setting, the input consists of corrupted expression matrices where zero or near-zero values arise due to technical dropouts, while the target outputs correspond to the original, uncorrupted gene expression values~\cite{wang2022imputation}. Ground truth data are obtained from real datasets where artificial dropout is introduced in a controlled manner~\cite{wang2022imputation,cheng2023evaluating,hou2020systematic}. Imputation models are trained by learning a mapping from the observed sparse data to the complete expression space, and performance is quantitatively evaluated using regression-based error metrics such as \gls{mse} and \gls{mae} computed at the gene, cell, or matrix level~\cite{cheng2023evaluating,dai2022scimc}.

\paragraph{Cell Clustering.} Cell clustering can be formulated as an unsupervised learning problem, where each cell is treated as an individual sample represented by a high-dimensional gene expression vector~\cite{kiselev2019challenges}. The objective is to group similar cells into clusters based on their expression profiles~\cite{kiselev2019challenges}. Since \gls{scrnaseq} data is inherently high-dimensional, with thousands of genes measured per cell, dimensionality reduction techniques, such as \gls{pca}~\cite{pearson1901liii}, \gls{t-sne}~\cite{maaten2008visualizing}, and \gls{umap}~\cite{mcinnes2020umap}, are often applied prior to clustering to reduce noise and improve computational efficiency~\cite{luecken2019current}. Cell clustering is typically performed as an initial downstream task and serves as a foundation for subsequent downstream tasks, such as marker gene analysis and cell type annotation~\cite{kiselev2019challenges,luecken2019current}. Meaningful clusters enable the identification of distinct cell populations and cellular states, whereas inaccurate clustering can lead to misleading biological interpretations~\cite{luecken2019current}. In this benchmark, dimensionality reduction is performed using \gls{pca}~\cite{pearson1901liii}, and the Leiden algorithm~\cite{traag2019louvain} is subsequently applied for cell clustering. The Leiden algorithm is a graph-based community detection algorithm that operates on a cell-cell similarity graph constructed from the \gls{scrnaseq} data and partitions cells into clusters by optimizing modularity, a quality function that measures the density of edges within clusters compared to edges between clusters~\cite{traag2019louvain}.

\paragraph{\texorpdfstring{\gls{de} Analysis}{DE Analysis}.} \gls{de} analysis can be formulated as a feature selection problem, in which the objective is to identify genes that exhibit statistically significant expression differences between conditions, such as disease versus healthy groups or treatment versus control groups~\cite{luecken2019current,scholtens2005analysis}. This is typically achieved by testing the null hypothesis that the expression levels of a given gene are identical between the 2 groups~\cite{love2014moderated}. \gls{de} analysis enables the identification of genes associated with specific biological processes or disease states and represents a core downstream task for linking gene expression changes to underlying biological phenomena~\cite{luecken2019current}. Accurate identification of \glspl{deg} is therefore crucial for understanding molecular mechanisms related to disease progression or treatment response~\cite{luecken2019current}. In this benchmark, 2 common \gls{de} analysis methods are used, namely MAST~\cite{finak2015mast}, which is a statistical framework using a hurdle model to account for the bimodal distribution of \gls{scrnaseq} data~\cite{finak2015mast,hou2020systematic}, and the Wilcoxon rank-sum test~\cite{mann1947test,wilcoxon1945individual,wolf2018scanpy}, in order to evaluate the performance and robustness of imputation methods across multiple \gls{de} analysis approaches~\cite{hou2020systematic}.

\paragraph{Marker Gene Analysis.} Marker gene analysis is an application of \gls{de} analysis, and can be formulated as a feature selection problem, in which the objective is to identify genes that best represent each cluster of cells~\cite{luecken2019current,pullin2024comparison}. This task is typically performed in 2 steps. First, \gls{de} analysis is conducted to identify genes whose expression levels significantly differ between a given cluster and the remaining clusters~\cite{luecken2019current,pullin2024comparison}. Second, genes are ranked based on \gls{lfc} or test statistics derived from \gls{de} analysis, and the top-ranked genes are selected as marker genes for each cluster~\cite{pullin2024comparison}. Marker gene analysis plays a crucial role in the interpretation of cell clusters, as marker genes provide insights into the biological functions and identities of distinct cell populations~\cite{pullin2024comparison,luecken2019current}. Accurate identification of marker genes enables reliable interpretation of the biological significance of cell clusters, and a deeper understanding of underlying cellular heterogeneity~\cite{pullin2024comparison,luecken2019current}. In this benchmark, rather than identifying marker genes for each cluster, the preservation of marker gene expression is evaluated by examing whether known marker genes of each cell type are expressed in the corresponding cell types of the imputed data.

\paragraph{Trajectory Analysis.} Trajectory analysis can be formulated as an unsupervised latent-structure inference task, where the objective is to infer continuous cellular progression and lineage relationships from \gls{scrnaseq} data~\cite{luecken2019current}. This ordering is commonly represented by pseudotime values assigned to each cell~\cite{luecken2019current}. Pseudotime is a continuous latent variable that captures the relative progression of cells through a biological process, such as differentiation or development~\cite{luecken2019current,haghverdi2016diffusion,wolf2019paga}. Pseudotime is typically inferred from cell-cell relationships by measuring distances between cells in the original expression space or in a reduced-dimensional representation~\cite{luecken2019current,haghverdi2016diffusion,wolf2019paga}. Trajectory analysis is essential for understanding dynamic cellular processes and identifying key regulatory genes involved in these processes~\cite{luecken2019current}. Reliable inference of pseudotime enables the discovery of temporal patterns of gene expression and provides insights into the mechanisms driving cellular transitions~\cite{luecken2019current}. In this benchmark, TSCAN~\cite{ji2016tscan} is used to perform trajectory inference. TSCAN first reduces the dimensionality of gene expression data using \gls{pca}, then clusters cells in the reduced space, and constructs a \gls{mst} connecting cluster centers to represent the trajectory structure~\cite{ji2016tscan}. Pseudotime values are subsequently assigned by projecting individual cells onto the nearest edge of the \gls{mst}~\cite{ji2016tscan}.

\paragraph{Cell Type Annotation.} Cell type annotation can be formulated as a supervised multi-class classification problem, in which inputs are gene expression vectors of each cell, and outputs are corresponding cell type labels~\cite{luecken2019current,pasquini2021automated,kiselev2019challenges}. The objective of this task is to learn a mapping from gene expression vectors to cell type labels based on known cell type labels~\cite{pasquini2021automated}. Cell types are predefined at different levels of granularity, such as broad cell types, e.g., T cells, B cells, and monocytes, or fine-grained cell subtypes, e.g., CD4$^+$ T cells, CD8$^+$ T cells, and regulatory T cells~\cite{pasquini2021automated,dominguezconde2022crosstissue}. A typical approach to perform cell type annotation is to compare the \gls{scrnaseq} data with previously annotated reference datasets using classification models~\cite{pasquini2021automated}. In this approach, a classifier is trained on a reference dataset to learn the mapping from gene expression vectors to cell type labels, and is subsequently used to predict cell type labels for new \gls{scrnaseq} data~\cite{pasquini2021automated}. This task is essential for the interpretation of \gls{scrnaseq} data, as it provides biological context for cells and clusters identified in the data~\cite{pasquini2021automated,kiselev2019challenges,lahnemann2020eleven}. Robust cell type annotation enables researchers to better understand cellular heterogeneity and the functional roles of different cell types in biological processes~\cite{luecken2019current,pasquini2021automated}. In this benchmark, scGPT~\cite{cui2024scgpt}, a foundation model for \gls{scrnaseq} data which supports cell type annotation, and \gls{1d-cnn} are used to evaluate the performance of cell type annotation across different imputation methods and \gls{scrnaseq} datasets. For scGPT, the pretrained model released by the authors\footnote{\url{https://github.com/bowang-lab/scGPT}}, which is trained on 33 million human cells, is used, and for \gls{1d-cnn}, the model is trained on the training set of each dataset.

\subsection{Evaluation Measures}
\label{subsec:method_evaluation_metrics}

All 16 imputation methods are evaluated across 6 downstream tasks that capture both numerical accuracy and biological relevance. Since each task serves a different analytical objective, a single metric is insufficient to fully characterize performance. Therefore, task-specific 18 different evaluation measures are utilized to assess reconstruction quality, clustering consistency, statistical agreement, temporal ordering, and classification accuracy. Together, these measures provide a comprehensive and fair comparison of all methods.

\paragraph{Numerical Gene Expression Recovery.} The evaluation of numerical gene expression recovery can be performed by directly comparing imputed gene expression values with ground truth values or by comparing with corresponding bulk RNA-seq data~\cite{cheng2023evaluating,hou2020systematic,dai2022scimc}. 4 distinct evaluation measures are used to directly evaluate the numerical gene expression recovery performance of all 16 imputation methods, namely \gls{mae}, \gls{medae}, \gls{lnd}, and \gls{mse}. Each measure is computed by comparing the imputed gene expression values with the ground truth expression values. \Gls{mae} is computed as the average of the absolute differences between imputed and ground truth values, which provides a more direct measure of average error~\cite{cheng2023evaluating}. \Gls{medae} is computed as the median of the absolute differences between imputed and ground truth values, which provides a similar measure as \gls{mae} without outliers~\cite{cheng2023evaluating}. \Gls{lnd} is calculated as the log-transformed difference between imputed and ground truth values, which allows assessment of over- or under-imputation~\cite{cheng2023evaluating}. \Gls{mse} is calculated as the average of the squared differences between imputed and ground truth values, which provides a measure of overall error magnitude~\cite{cheng2023evaluating,dai2022scimc}. In addition to comparison with ground truth values, comparison with matched bulk RNA-seq data is performed to evaluate the performance of imputation methods in recovering gene expression patterns~\cite{hou2020systematic} with 2 distinct evaluation measures, namely \gls{pcc} and \gls{mcc}. \Gls{pcc} measures the correlation between pseudo-bulk expression values, calculated by averaging imputed gene expression values across all cells, and bulk RNA-seq expression values~\cite{hou2020systematic}. \Gls{mcc} measures the median correlation between imputed gene expression values of individual cells and bulk RNA-seq expression values~\cite{hou2020systematic}. Both measures are calculated using \gls{scc}~\cite{spearman1904proof}.
\begin{gather}
    \text{MAE} = \frac{1}{N} \sum_{i=1}^{N} |\hat{y}_i - y_i| \\[6pt]
    \text{MedAE} = \mathrm{median}\{|\hat{y}_i - y_i|\}_{i=1}^N \\[6pt]
    \text{LND} =
    \begin{cases}
        \log_{2}(\hat{y}_i - y_i + 1) & \text{if } \hat{y}_i \geq y_i \\
        -\log_{2}(- \hat{y}_i + y_i + 1) & \text{if } \hat{y}_i < y_i
    \end{cases} \\[6pt]
    \text{MSE} = \frac{1}{N} \sum_{i=1}^{N} (\hat{y}_i - y_i)^2 \\[6pt]
    \text{PCC} = \text{SCC}(\hat{y}_{\text{pseudo-bulk}}, y_{\text{bulk}}) \\[6pt]
    \text{MCC} = \text{median} \left\{ \text{SCC}(\hat{y}_{i}, y_{\text{bulk}}) \right\}_{i=1}^{N}
\end{gather}
Here, $N$ is the total number of imputed entries, $\hat{y}_i$ is the imputed expression value of the $i$-th entry, and $y_i$ is the corresponding ground truth expression value. $\hat{y}_{\text{pseudo-bulk}}$ is the pseudo-bulk expression values calculated by averaging imputed gene expression values across all cells, and $y_{\text{bulk}}$ is the matched bulk RNA-seq expression values~\cite{hou2020systematic}. \Gls{scc} is calculated as $\text{SCC(X, Y)} = 1 - \frac{6 \sum_{i=1}^{N} D^2}{N(N^2 - 1)}$, where $N$ is the total number of genes, and $D$ is the difference between the ranks of the given 2 variables $X$ and $Y$~\cite{hou2020systematic}.

\paragraph{Cell Clustering.} 4 distinct evaluation measures are used to evaluate cell clustering performance, namely \gls{ari}~\cite{hubert1985comparing}, \gls{nmi}, purity, and \gls{sc}~\cite{cheng2023evaluating,dai2022scimc,hou2020systematic}. \Gls{ari} measures the similarity between 2 clustering results by considering all pairs of cells and counting pairs that are assigned in the same or different clusters in the clustering results of imputed data and ground truth data~\cite{cheng2023evaluating,dai2022scimc,hou2020systematic}. \Gls{nmi} measures the mutual dependence between the clustering results of imputed data and ground truth data~\cite{cheng2023evaluating}. Purity measures clustering quality by quantifying how homogeneous each predicted cluster is with respect to ground truth labels~\cite{cheng2023evaluating}. \Gls{sc} measures clustering quality by quantifying cohesion within clusters and separation between clusters~\cite{cheng2023evaluating,dai2022scimc}.
\begin{gather}
    \text{ARI} = \frac{\sum_{ij} \binom{n_{ij}}{2} - E}{\frac{1}{2} \left[ \sum_i \binom{a_i}{2} + \sum_j \binom{b_j}{2} \right] - E} \\*[6pt]
    E = \left[ \sum_i \binom{a_i}{2} \sum_j \binom{b_j}{2} \right] \Big/ \binom{n}{2} \nonumber \\[6pt]
    \text{NMI} = \frac{I(X; Y)}{\sqrt{H(X) H(Y)}} \\[6pt]
    \text{Purity} = \frac{1}{n} \sum_{k=1}^{K} \max_{j} |c_k \cap t_j| \\[6pt]
    \text{SC} = \frac{b(i) - a(i)}{\max\{a(i), b(i)\}}
\end{gather}
Here, for \gls{ari}, $n$ is the total number of cells, $i$ and $j$ are cluster indices in the clustering results of imputed data and ground truth data, respectively, $n_{ij}$ is the number of cells in both cluster $i$ and cluster $j$, $a_i = \sum_j n_{ij}$, $b_j = \sum_i n_{ij}$, and $E$ is the expected index under random cluster assignment. For \gls{nmi}, $X$ and $Y$ are the cluster assignments from the clustering results of imputed data and ground truth data, respectively, $I(X; Y)$ is the mutual information between $X$ and $Y$, and $H(X)$ and $H(Y)$ are the entropies of $X$ and $Y$, respectively. For purity, $n$ is the total number of cells, $K$ is the number of predicted clusters, $c_k$ is the set of cells in predicted cluster $k$, and $t_j$ is the set of cells in ground truth cluster $j$. For \gls{sc}, $a(i)$ is the average distance between cell $i$ and all other cells in the same cluster, and $b(i)$ is the minimum average distance between cell $i$ and all cells in other clusters.

\paragraph{\texorpdfstring{\gls{de} Analysis}{DE Analysis}.} 2 distinct evaluation measures are used to evaluate \gls{de} analysis performance, namely \gls{iou} and \gls{fpdeg}. \gls{iou} measures the overlap between the sets of genes in the 2 different groups, e.g., \glspl{deg} identified from imputed \gls{scrnaseq} data and from the corresponding bulk RNA-seq data. \Gls{fpdeg} measures the number of genes identified as \glspl{deg} that are not true \glspl{deg}.
\begin{gather}
    \mathrm{IoU} = \frac{|G_\mathrm{A} \cap G_\mathrm{B}|}{|G_\mathrm{A} \cup G_\mathrm{B}|} \\[6pt]
    \mathrm{FPDEG} = |G_\mathrm{null}|
\end{gather}
Here, $G_\mathrm{A}$ and $G_\mathrm{B}$ are the sets of genes in groups A and B, respectively, and $G_\mathrm{null}$ is the set of \glspl{deg} identified between control and target groups randomly sampled from the same cell population.

\paragraph{Marker Gene Analysis.} The evaluation is done qualitatively through visual inspection of marker gene expression distributions and cell type separation. Violin plots are used to compare the distribution of known marker gene expression levels across cell types, which assess whether imputed data retain expected cell-type-specific enrichment patterns. In addition, \gls{umap} visualizations are used to evaluate whether imputed data produce clear separation of distinct cell types in low-dimensional space. Heatmaps of marker gene expression values across cell types further complement this evaluation by illustrating whether imputation methods recover distinct expression signatures for each cell type. Together, these visualizations assess the degree to which each imputation method preserves the biological signal encoded in established marker genes.

\paragraph{Trajectory Analysis.} 2 distinct evaluation metrics are used to evaluate trajectory analysis performance, namely \gls{pos}~\cite{ji2016tscan,dai2022scimc} and \gls{krcc}~\cite{kendall1938new,dai2022scimc}. \Gls{pos} is calculated by summing scores that characterize how well the inferred cell ordering matches the expected ordering based on external information. \Gls{krcc} is computed to measure the correlation between the inferred pseudotime values and the true cell development labels.
\begin{gather}
    \text{POS} = \sum_{i=1}^{n-1} \sum_{j>i}^{n} g(i, j) \\[6pt]
    \text{KRCC} = \frac{4C}{n(n-1)} - 1
\end{gather}
Here, $n$ is the number of cells, and $g(i, j)$ is a score that characterizes how well the order of the $i$-th and $j$-th cells in the ordered path matches their expected order based on the external information~\cite{dai2022scimc,ji2016tscan}, and $C$ is the number of concordant pairs~\cite{kendall1938new,dai2022scimc}. See \citet{ji2016tscan} for a detailed definition of $g(i, j)$.

\paragraph{Cell Type Annotation.} 4 distinct evaluation measures are used to evaluate cell type annotation performance, namely \gls{acc}, \gls{pr}, \gls{rc}, and \gls{f1}. Each measure is computed using a macro-averaging approach to ensure equal weighting for all cell types irrespective of their numbers of cells. \Gls{acc} is calculated as the average of individual accuracy scores across all cell types. For a single cell type, accuracy is computed as the ratio of correctly predicted samples, i.e., true positives and true negatives, to the total samples. \Gls{pr} is calculated as the average of the ratio of true positives to total predicted positives for each cell type, with the PR of a single cell type computed as true positives divided by predicted positives. \Gls{rc} is the average of the ratio of true positives to total actual positives for each cell type, with the \gls{rc} of a single cell type computed as true positives divided by actual positives. \Gls{f1} is the harmonic mean of \gls{pr} and \gls{rc}, averaged across all cell types. For individual cell types, the \gls{f1} is computed as the harmonic mean of that cell type's \gls{pr} and \gls{rc}.
\begin{gather}
    \text{ACC} = \frac{1}{K} \sum_{k=1}^{K} \frac{\mathrm{TP}_k + \mathrm{TN}_k}{\mathrm{TP}_k + \mathrm{TN}_k + \mathrm{FP}_k + \mathrm{FN}_k} \\[6pt]
    \text{PR} = \frac{1}{K} \sum_{k=1}^{K} \frac{\mathrm{TP}_k}{\mathrm{TP}_k + \mathrm{FP}_k} \\[6pt]
    \text{RC} = \frac{1}{K} \sum_{k=1}^{K} \frac{\mathrm{TP}_k}{\mathrm{TP}_k + \mathrm{FN}_k} \\[6pt]
    \text{F1} = \frac{1}{K} \sum_{k=1}^{K} \frac{2 \mathrm{PR}_k \mathrm{RC}_k}{\mathrm{PR}_k + \mathrm{RC}_k}
\end{gather}
Here, $K$ is the number of cell types, and for each cell type $k$, $\mathrm{TP}_k$, $\mathrm{TN}_k$, $\mathrm{FP}_k$, and $\mathrm{FN}_k$ denote the numbers of true positive, true negative, false positive, and false negative cell type annotations compared to ground truth cell type annotations, respectively.

\subsection{Experimental Setup}
\label{subsec:method_experimental_setup}

Our benchmarking framework is implemented using Python and R. After collecting datasets, each real dataset is formatted into a \href{https://anndata.readthedocs.io}{\texttt{anndata}}~\cite{virshup2024anndata} object, which is a widely used sparse matrix format for \gls{scrnaseq} data in Python. Simulated datasets are generated using the \href{https://www.bioconductor.org/packages/release/bioc/html/splatter.html}{\texttt{Splatter}}~\cite{zappia2017splatter} package in R. Data imputation methods are run with the code or package released by the method authors, or with our Python ports that follow the original code, and with the default parameters of their original implementations except for scIDPMs~\cite{zhang2025scidpms}. Supplementary Table S7 lists the implementation, the parameters, the cells used for training, and the random seed of each method. For scIDPMs, the model size, the number of diffusion steps, and the number of samples drawn at inference are reduced, because running it with its original settings on all datasets would exceed our computational resources. We denote this configuration as scIDPMs$^\dagger$ throughout. Downstream tasks are performed using the \href{https://scanpy.readthedocs.io}{\texttt{Scanpy}}~\cite{wolf2018scanpy} package in Python, except for \gls{de} analysis which is performed using the \href{https://github.com/RGLab/MAST}{\texttt{MAST}}~\cite{finak2015mast} and \href{https://bioconductor.org/packages/release/bioc/html/limma.html}{\texttt{limma}}~\cite{ritchie2015limma} package in R, trajectory analysis which is performed using the \href{https://github.com/zji90/TSCAN}{\texttt{TSCAN}}~\cite{ji2016tscan} package in R, and cell type annotation which is performed using the \href{https://github.com/bowang-lab/scGPT}{\texttt{scGPT}}~\cite{cui2024scgpt} package in Python and \gls{1d-cnn} implemented with \href{https://pytorch.org}{\texttt{PyTorch}}~\cite{ansel2024pytorch} package in Python. The evaluation metrics are calculated on top of the \href{https://scikit-learn.org}{\texttt{scikit-learn}}~\cite{pedregosa2011scikitlearn} package in Python. All visualizations are created using the \href{https://ggplot2.tidyverse.org}{\texttt{ggplot2}}~\cite{wickham2016ggplot2} package in R.

\section{Results}
\label{sec:results}

\subsection{Numerical Gene Expression Recovery}
\label{subsec:results_numerical_recovery}

\begin{figure*}[p]
    \centering
    \includegraphics[width=\textwidth,height=0.92\textheight,keepaspectratio]{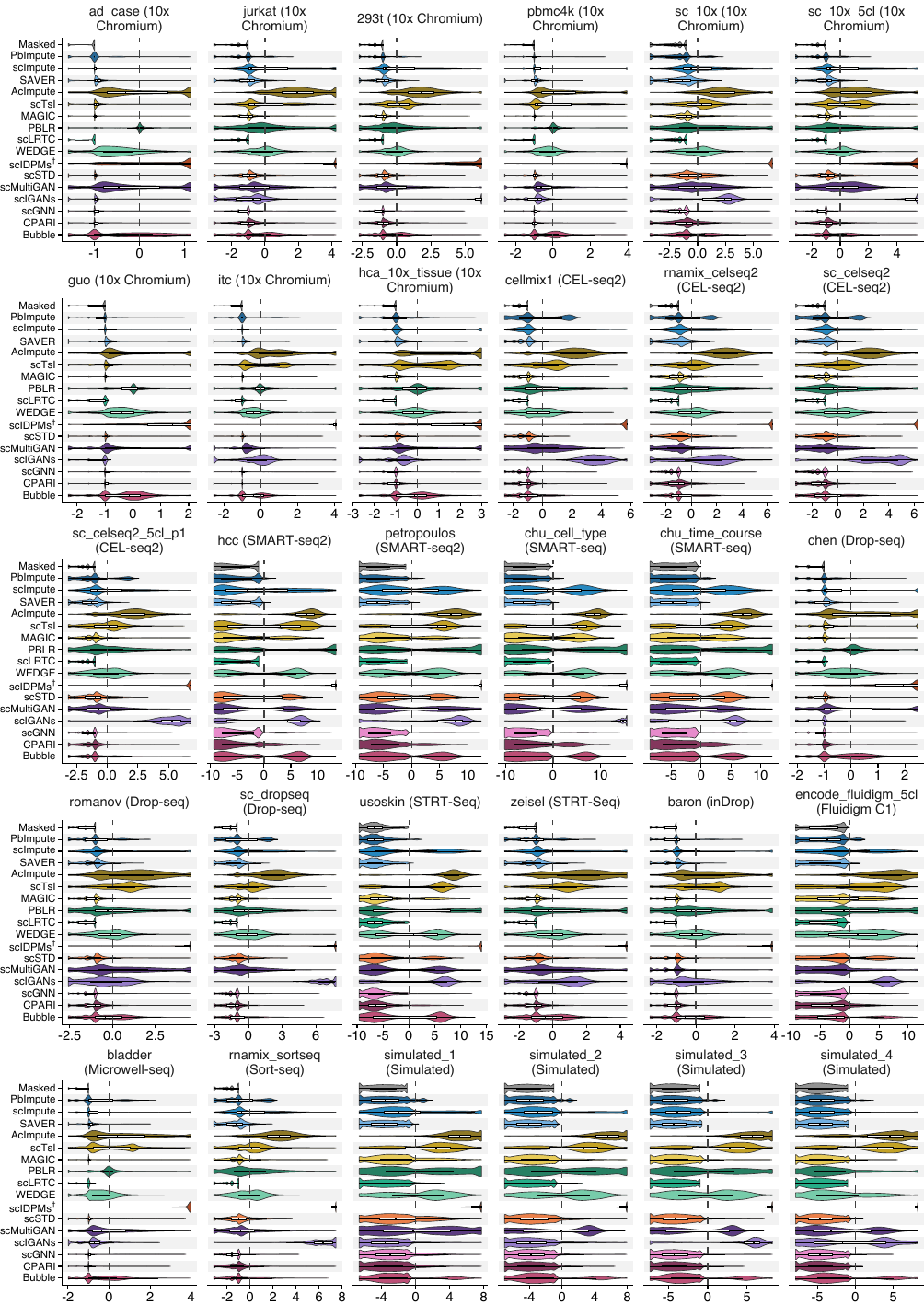}
    \caption{Distribution of LND between imputed and ground truth expression values for each imputation method. The x-axis represents LND values, and the y-axis represents different imputation methods. $^\dagger$Run with reduced settings (see \cref{subsec:method_experimental_setup}).}
    \label{fig:results_numerical_recovery_ground_truth_lnd}
\end{figure*}

\begin{figure*}[p]
    \centering
    \includegraphics[width=\textwidth,height=0.82\textheight,keepaspectratio]{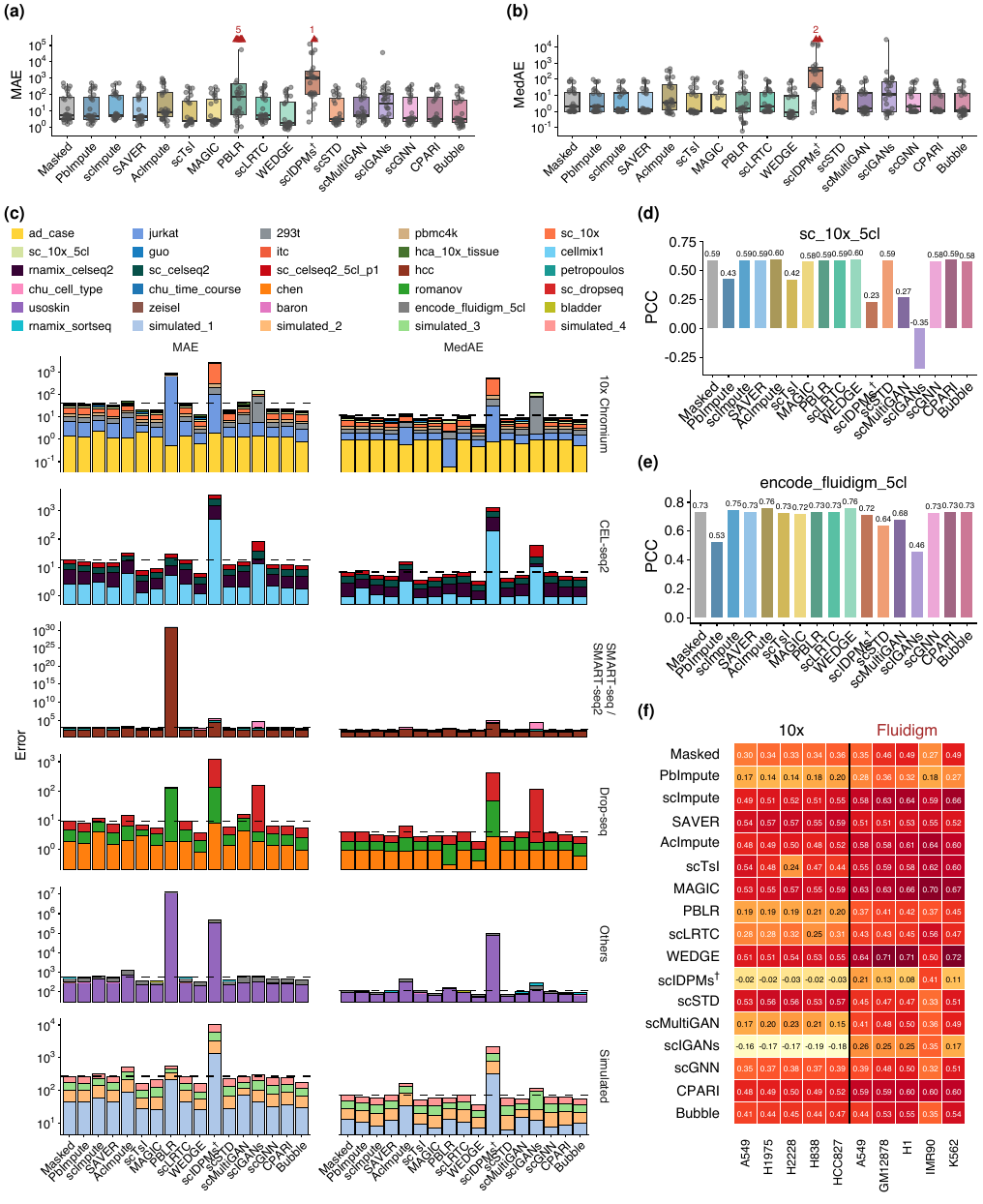}
    \caption{Numerical gene expression recovery performance. \textbf{(a)}--\textbf{(b)} MAE and MedAE, respectively. The x-axis represents different imputation methods, and the y-axis represents error values in a log scale. Values above the upper end of the highest whisker are shown as red triangles at the top of each panel, with their number. \textbf{(c)} Protocol-wise total MAE and MedAE. The x-axis represents different imputation methods, and the y-axis represents total error values in a log scale. \textbf{(d)}--\textbf{(e)} PCC. The x-axis represents different imputation methods, and the y-axis represents PCC values. \textbf{(f)} MCC. The x-axis represents different cell lines, and the y-axis represents different imputation methods. $^\dagger$Run with reduced settings (see \cref{subsec:method_experimental_setup}).}
    \label{fig:results_numerical_recovery_ground_truth}
\end{figure*}

\cref{fig:results_numerical_recovery_ground_truth_lnd} represents the distribution of \gls{lnd} values between imputed and ground truth expression values for 16 imputation methods in terms of 26 real and 4 simulated datasets. The width of each violin plot represents the density of \gls{lnd} values, and the box plot shows the median and \gls{iqr} of \gls{lnd} values. $\mathrm{LND} = 0$ indicates that all imputed expression values are identical to the ground truth values, $\mathrm{LND} > 0$ indicates over-imputation, where the imputed values are greater than the ground truth values, and $\mathrm{LND} < 0$ indicates under-imputation, i.e., the imputed values are less than the ground truth values.

The comparison of \gls{lnd} distributions across 16 imputation methods in terms of 30 different datasets shows that WEDGE achieves the best overall performance, with medians of \gls{lnd} values closest to zero in 14 of the 30 datasets, followed by PBLR in 13 datasets. Conversely, scIDPMs$^\dagger$ shows over-imputation in all 30 datasets. scLRTC exhibits under-imputation in all datasets, with median \gls{lnd} values identical to those of the masked baseline in 23 datasets. The remaining 12 methods, including scMultiGAN, scTsI, Bubble, scSTD, MAGIC, SAVER, AcImpute, scImpute, scIGANs, CPARI, PbImpute, and scGNN, show moderate performance. 

Protocol-wise analysis of \gls{lnd} distributions is essential because protocols differ in sparsity, noise, and dropout characteristics, which directly affect imputation behavior. Across 2 protocols, namely CEL-seq2 and Drop-seq, WEDGE maintains \gls{lnd} values closest to zero, which indicates stable reconstruction. In contrast, scIDPMs$^\dagger$ exhibits positive \gls{lnd} shifts in all protocols, which reflect systematic over-imputation. Conversely, scLRTC consistently demonstrates under-imputation in all protocols. Protocols based on full-length sequencing, including SMART-seq, SMART-seq2, and Fluidigm C1, show greater variability overall. All methods display broader distributions, and none achieve median \gls{lnd} values close to zero. These patterns suggest increased difficulty for accurate imputation. The remaining methods exhibit intermediate, dataset-dependent behavior with moderate deviations around zero. Collectively, these findings show that the performance of WEDGE depends on the protocol, while revealing distinct protocol-dependent biases for competing methods.

\namecrefs{fig:results_numerical_recovery_ground_truth}~\labelcref{fig:results_numerical_recovery_ground_truth}(a) and (b) represent the box plots for \gls{mae} and \gls{medae} between imputed and ground truth expression values for 16 imputation methods in terms of 26 real and 4 simulated datasets. \gls{mae} shows the overall performance of the methods considering outliers, while \gls{medae} shows the overall performance of the methods without outliers. Lower \gls{mae} and \gls{medae} values indicate better recovery performance, as they suggest that the imputed expression values are closer to the ground truth values.

A thorough analysis of \gls{mae} and \gls{medae} among the 16 imputation methods reveals that WEDGE exhibits the lowest \gls{mae} and \gls{medae}. In addition, 14 methods, namely PbImpute, scImpute, SAVER, AcImpute, scTsI, MAGIC, PBLR, scLRTC, scSTD, scMultiGAN, scIGANs, scGNN, CPARI, and Bubble, show moderate results. On the other hand, scIDPMs$^\dagger$ shows the worst \gls{mae} and \gls{medae}.

\cref{fig:results_numerical_recovery_ground_truth}(c) shows total \gls{mae} and total \gls{medae} across all datasets for each imputation method. WEDGE shows the best \gls{mae} across 5 protocols, namely 10x Chromium, CEL-seq2, SMART-seq2, SMART-seq, and Drop-seq. Similarly, WEDGE shows the best \gls{medae} across 3 protocols, namely 10x Chromium, CEL-seq2, and Drop-seq. On the other hand, scIDPMs$^\dagger$ exhibits the highest \gls{medae} across the 5 protocols, namely 10x Chromium, CEL-seq2, SMART-seq, SMART-seq2, and Drop-seq, and the highest \gls{mae} across 3 protocols, namely 10x Chromium, CEL-seq2, and Drop-seq.

\namecrefs{fig:results_numerical_recovery_ground_truth}~\labelcref{fig:results_numerical_recovery_ground_truth}(d) and (e) show \gls{pcc} between pseudo-bulk and the corresponding bulk RNA-seq data for 16 imputation methods in terms of 2 cell line datasets, namely sc\_10x\_5cl and encode\_fluidigm\_5cl. A higher \gls{pcc} indicates that pseudo-bulk data is highly correlated with the corresponding bulk RNA-seq data. The comparison of \gls{pcc} across the 16 methods shows that AcImpute and WEDGE achieve the best performance. In addition, 13 methods, namely PbImpute, scImpute, SAVER, scTsI, MAGIC, PBLR, scLRTC, scIDPMs$^\dagger$, scSTD, scMultiGAN, scGNN, CPARI, and Bubble, show moderate performance. On the other hand, scIGANs shows the worst performance.

\cref{fig:results_numerical_recovery_ground_truth}(f) shows \gls{mcc} between imputed \gls{scrnaseq} data and the corresponding bulk RNA-seq data at the cell line level for 16 imputation methods in terms of 2 cell line datasets, namely sc\_10x\_5cl and encode\_fluidigm\_5cl. A higher \gls{mcc} indicates that imputed \gls{scrnaseq} data is highly correlated with the corresponding bulk RNA-seq data. The comparison of \gls{mcc} across the 16 methods shows that MAGIC and WEDGE achieve the best performance. In addition, 12 methods, namely PbImpute, scImpute, SAVER, AcImpute, scTsI, PBLR, scLRTC, scSTD, scMultiGAN, scGNN, CPARI, and Bubble, show moderate performance. On the other hand, scIDPMs$^\dagger$ and scIGANs show the worst performance.

In summary, for comparison with ground truth data, WEDGE shows the best overall performance, while scIDPMs$^\dagger$ shows the worst performance. Furthermore, imputation quality is protocol dependent, as methods show largely consistent behavior on 10x Chromium, CEL-seq2, and Drop-seq datasets, whereas SMART-seq, SMART-seq2, and Fluidigm C1 datasets demonstrate higher instability, characterized by systematic over- or under-imputation across the 16 imputation methods. These methods tend to struggle with SMART-seq, SMART-seq2, and Fluidigm C1 datasets because these datasets are generated only using read-counts, whereas other datasets use \glspl{umi}~\cite{cheng2023evaluating}. The absence of \glspl{umi} can lead to duplicate read counts in the \gls{scrnaseq} data, which results in increased technical noise in the data~\cite{cheng2023evaluating}. For comparison with bulk RNA-seq data, WEDGE achieves the best overall performance. On the other hand, scIGANs shows the worst correlation with bulk RNA-seq data at both the pseudo-bulk and cell line levels, despite its moderate numerical recovery of ground truth data. This suggests that comparable ground truth recovery does not necessarily translate to faithful agreement with bulk RNA-seq data.

\subsection{Cell Clustering}
\label{subsec:results_cell_clustering}

\begin{figure*}[p]
    \centering
    \includegraphics[width=\textwidth,height=0.92\textheight,keepaspectratio]{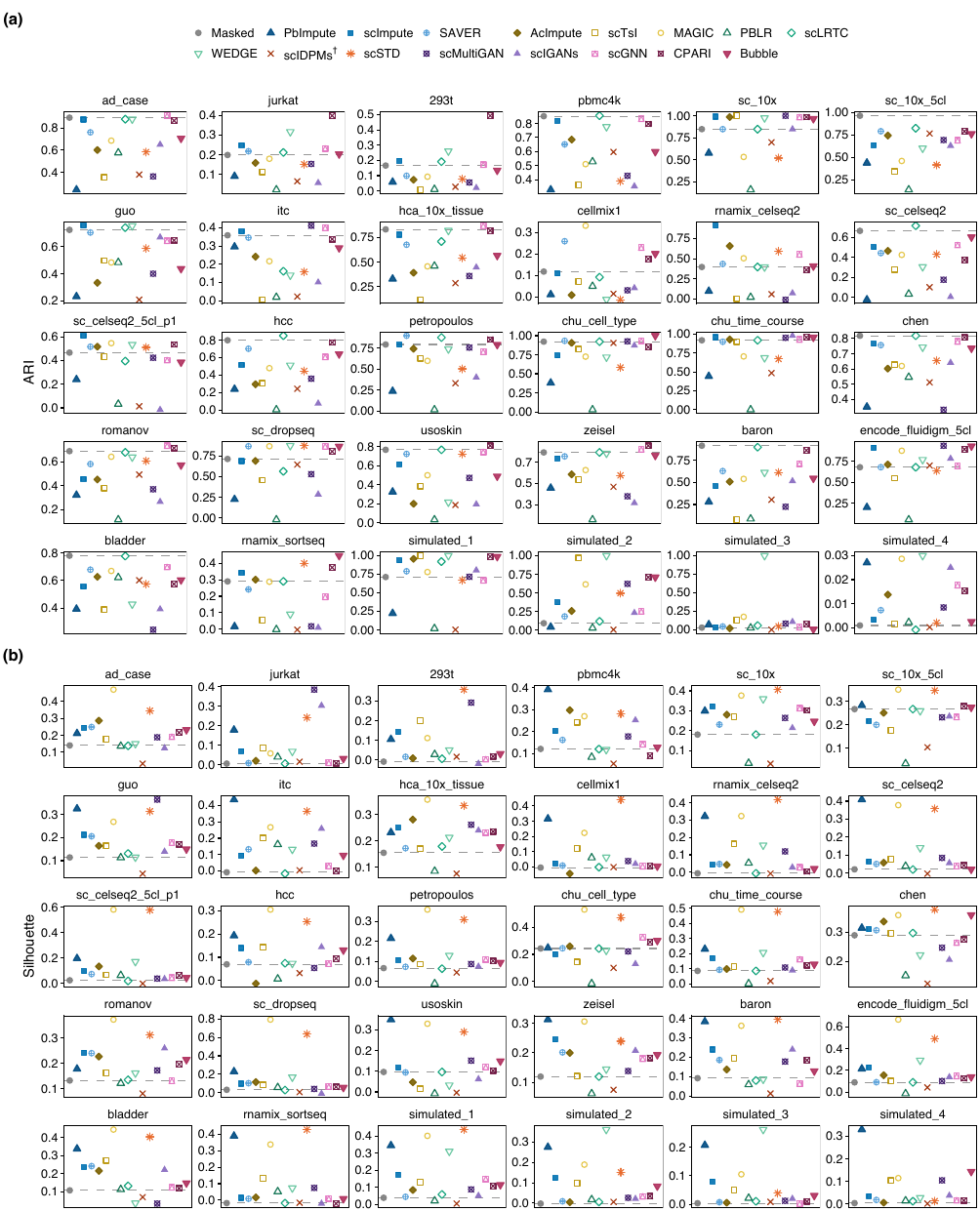}
    \caption{Cell clustering consistency and coherency performance. \textbf{(a)} ARI. Each plot shows ARI scores of different imputation methods. Each point in a plot represents different methods. The horizontal dashed line represents the masked baseline value. \textbf{(b)} SC. Each plot shows SC scores of different imputation methods. Each point in a plot represents different methods. The horizontal dashed line represents the masked baseline value. $^\dagger$Run with reduced settings (see \cref{subsec:method_experimental_setup}).}
    \label{fig:cell_clustering_multiplot}
\end{figure*}

\cref{fig:cell_clustering_multiplot}(a) represents \gls{ari} of cell clustering based on the imputed and ground truth data for the 16 imputation methods in terms of 26 real and 4 simulated datasets. Particularly, it demonstrates the consistency between cell clustering using imputed and ground truth data. $\mathrm{ARI} = 1$ shows clusters match perfectly, $\mathrm{ARI} = 0$ shows cell clustering performance is equivalent to randomly assigning clusters, and $\mathrm{ARI} < 0$ shows cell clustering performance is worse than randomly assigning clusters. Out of 16 distinct imputation methods, scLRTC exhibits the highest \gls{ari} scores in 7 datasets. In addition, 14 methods, namely PbImpute, scImpute, SAVER, AcImpute, scTsI, MAGIC, WEDGE, scIDPMs$^\dagger$, scSTD, scMultiGAN, scIGANs, scGNN, CPARI, and Bubble, show moderate \gls{ari} scores. On the other hand, PBLR shows the lowest \gls{ari} scores for 12 datasets. Moreover, for 3 datasets, including baron, bladder, and sc\_10x\_5cl, none of the 16 methods exceed the \gls{ari} scores of the masked baseline, and for 25 of the 30 datasets, at least half of the 16 methods do not exceed it. This indicates that imputation does not necessarily improve cell clustering and can even degrade performance compared to using the masked baseline data.

\cref{fig:cell_clustering_multiplot}(b) represents \gls{sc} for the 16 imputation methods in terms of 26 real and 4 simulated datasets. $\mathrm{SC} = 1$ represents cells are perfectly assigned to highly dense and isolated clusters, $\mathrm{SC} = 0$ represents cells are located at the boundaries between 2 clusters, and $\mathrm{SC} < 0$ represents cells are inaccurately assigned to clusters. The comparison of \gls{sc} scores among the 16 methods shows that MAGIC achieves the best performance for 12 datasets. On the other hand, scIDPMs$^\dagger$ shows the worst \gls{sc} scores for 13 datasets. The remaining 14 methods, namely PbImpute, scImpute, SAVER, AcImpute, scTsI, PBLR, scLRTC, WEDGE, scSTD, scMultiGAN, scIGANs, scGNN, CPARI, and Bubble, show moderate \gls{sc} scores.

\begin{figure*}[p]
    \centering
    \includegraphics[width=\textwidth,height=0.92\textheight,keepaspectratio]{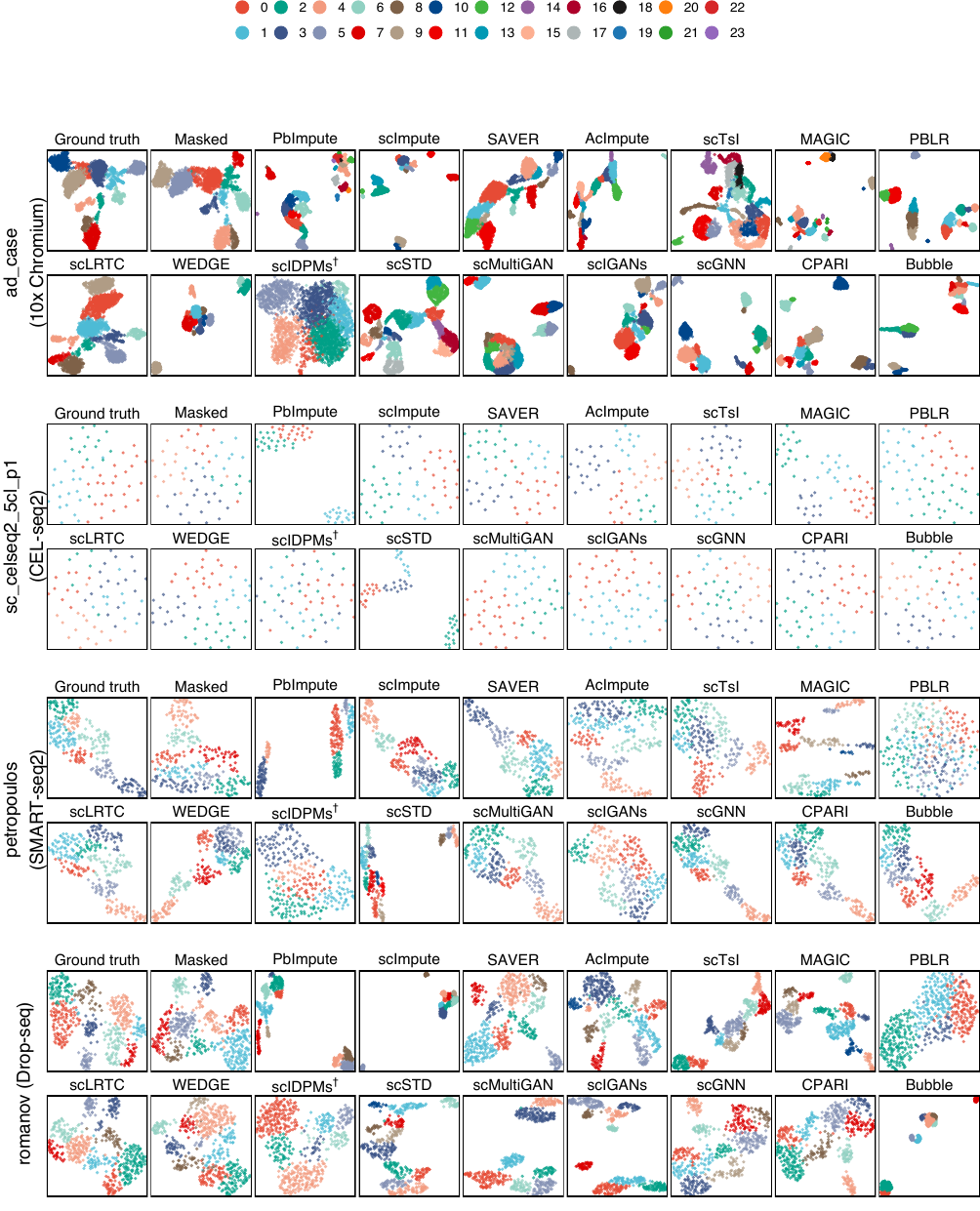}
    \caption{UMAP visualization of the cell clustering using 4 real datasets with different protocols. Each plot shows the UMAP visualization of the method. Different colors represent different clusters. $^\dagger$Run with reduced settings (see \cref{subsec:method_experimental_setup}).}
    \label{fig:cell_clustering_umap_real}
\end{figure*}

\begin{figure*}[p]
    \centering
    \includegraphics[width=\textwidth,height=0.92\textheight,keepaspectratio]{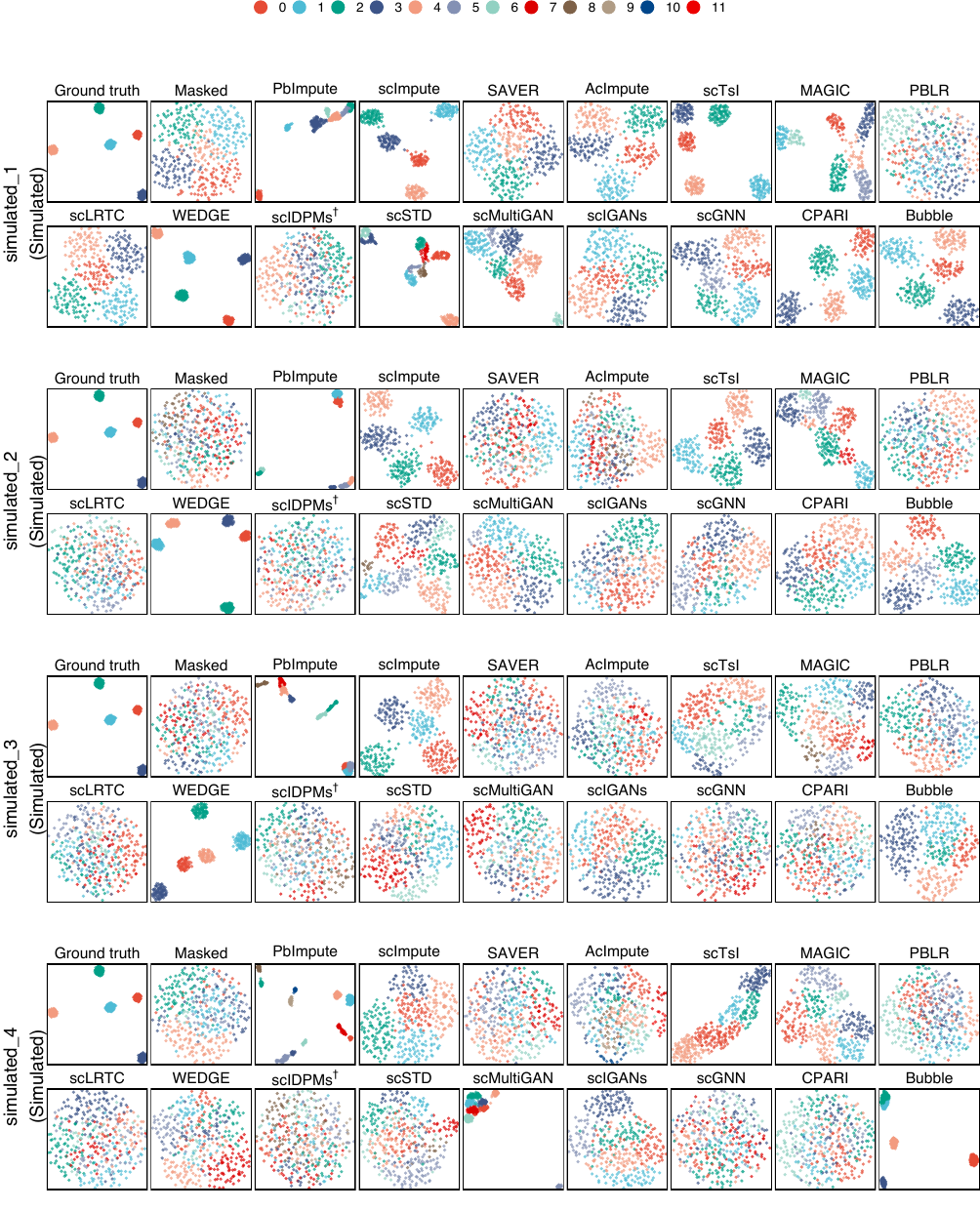}
    \caption{UMAP visualization of the cell clustering using 4 simulated datasets with different dropout rates. Each plot shows the UMAP visualization of the method. Different colors represent different clusters. $^\dagger$Run with reduced settings (see \cref{subsec:method_experimental_setup}).}
    \label{fig:cell_clustering_umap_simulated}
\end{figure*}

\cref{fig:cell_clustering_umap_real,fig:cell_clustering_umap_simulated} represent \gls{umap} visualization of the cell clustering for the 16 imputation methods in terms of 4 real datasets, and 4 simulated datasets, respectively. 4 datasets are selected from 4 protocols, including 10x Chromium, CEL-seq2, SMART-seq2, and Drop-seq, to perform a qualitative evaluation. A qualitative assessment of \gls{umap} plots provides a visual perspective on cluster consistency and coherency that complements the quantitative evaluations. The comparison of cluster structures among the 16 methods shows that MAGIC and scSTD produce the most visually coherent clusters across 4 real datasets, including sc\_celseq2\_5cl\_p1 with small samples. On the other hand, PBLR and scIDPMs$^\dagger$ show the least coherent cluster structures across 4 real datasets, with clusters appearing fragmented or poorly separated compared to the cell clustering based on the ground truth data. In the simulated datasets, WEDGE maintains visually distinct clusters across 3 simulated datasets, namely simulated\_1, simulated\_2, and simulated\_3. On the other hand, the remaining 15 methods show limited ability to recover the cluster structure of the ground truth data. This suggests that these 15 methods have lower robustness to dropout-induced sparsity.

\input{tables/nmi}
\input{tables/purity}

\cref{tab:nmi,tab:purity} report \gls{nmi} and purity scores, respectively, comparing cell clustering results from the imputed data with those from ground truth data across the 16 imputation methods for the 26 real and 4 simulated datasets. An \gls{nmi} score of $1$ indicates perfect agreement between the 2 clustering results, whereas $0$ indicates independence. Similarly, a purity score of $1$ indicates that each predicted cluster contains cells from a single ground truth cluster, while $0$ indicates complete mixing of cells from different ground truth clusters. The analyses of \gls{nmi} and purity scores among the 16 imputation methods show that these scores are largely consistent with the \gls{ari} scores, where scLRTC and CPARI exhibit the best \gls{nmi} and purity scores, respectively, in 8 datasets, PBLR shows the lowest \gls{nmi} and purity scores in 13 and 15 datasets, respectively, and the remaining methods show moderate \gls{nmi} and purity scores. This consistency across multiple cell clustering metrics reinforces the robustness of the observed performance differences among the 16 methods.

In summary, both quantitative and qualitative evaluations reveal substantial variability in cell clustering performance across the 16 imputation methods in terms of 26 real and 4 simulated datasets. scLRTC shows the best consistency performance, as supported by \gls{ari} and \gls{nmi}, and MAGIC shows the best coherency performance, as supported by \gls{sc} and \gls{umap} visualization. Conversely, PBLR shows the worst consistency performance, and PBLR and scIDPMs$^\dagger$ show the worst coherency performance.

\subsection{\texorpdfstring{\Gls{de}}{DE} Analysis}

\begin{figure*}[tb]
    \centering
    \includegraphics[width=\textwidth,keepaspectratio]{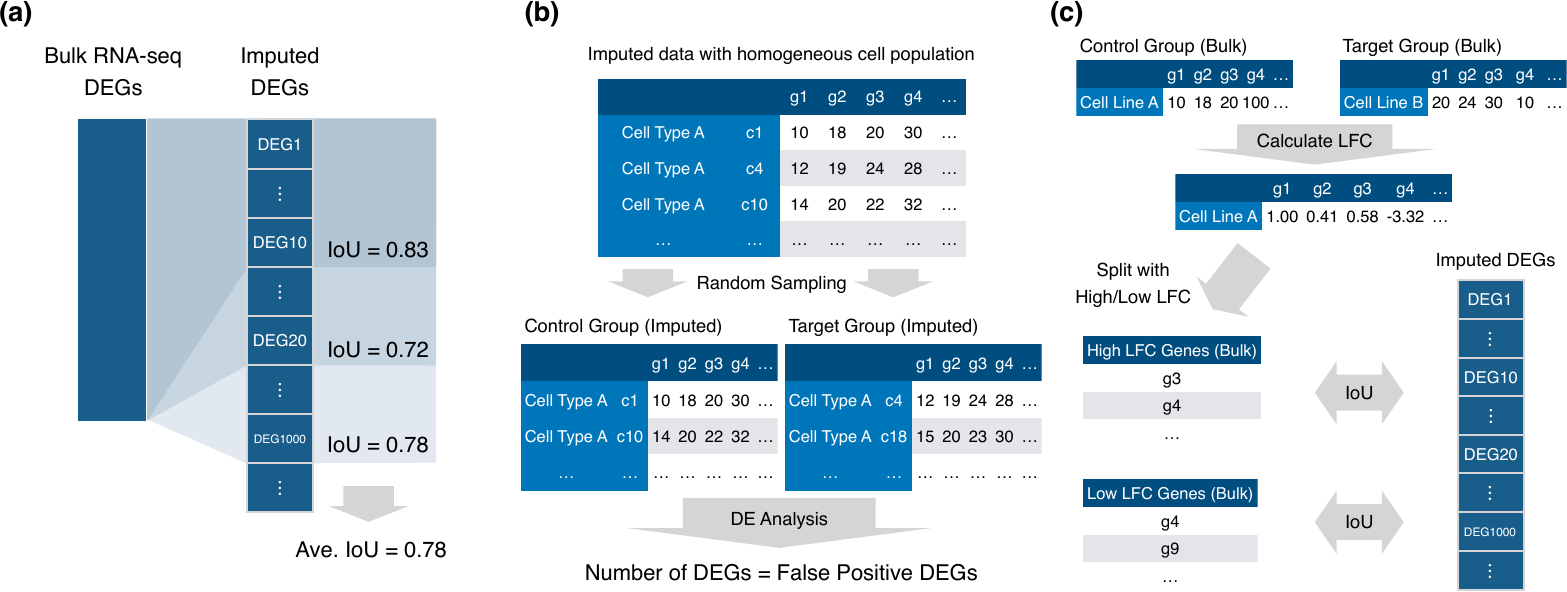}
    \caption{The overview of 3 complementary analyses of DE analysis. \textbf{(a)} DE enrichment analysis, \textbf{(b)} null DE analysis, and \textbf{(c)} effect size analysis.}
    \label{fig:de_analysis_diagram}
\end{figure*}

Following \citet{hou2020systematic}, 3 different \gls{de} analyses are conducted, namely \gls{de} enrichment analysis, null \gls{de} analysis, and effect size analysis, as illustrated in \cref{fig:de_analysis_diagram}. The \gls{de} enrichment analysis evaluates how well \glspl{deg} from imputed \gls{scrnaseq} data recover \glspl{deg} identified from bulk RNA-seq data, which serve as the ground truth \glspl{deg}. The \glspl{deg} from imputed \gls{scrnaseq} data are ranked by $p$ values or \gls{lfc} if there is a tie for $p$ values, and the \gls{iou} between the bulk RNA-seq \glspl{deg} and the top $10i$ imputed \glspl{deg} is computed for $i$ from 1 to 100. The average \gls{iou} across all 100 values of $i$ is used to measure the performance. The null \gls{de} analysis assesses the robustness of imputed data to false positive \glspl{deg}. Ideally, \glspl{deg} should not be identified when control and target groups belong to the same cell population, and any identified \glspl{deg} can be treated as false positive \glspl{deg} under such conditions. The imputed \gls{scrnaseq} data is filtered to a single cell type or cell line to obtain a homogeneous cell population. From this cell population, $N_1$ and $N_2$ cells are randomly sampled as control and target groups, respectively, where $N_1 \leq N_2$ and $N_1, N_2 \in \{10, 50, 100\}$. All 6 combinations, namely $(N_1, N_2) = (10,10)$, $(10,50)$, $(10, 100)$, $(50, 50)$, $(50, 100)$, $(100, 100)$, are tested to assess robustness across varying sample sizes and group balances, and \gls{de} analysis is performed between the control and target groups. For encode\_fluidigm\_5cl, only $(N_1, N_2) = (10, 10)$ is tested because its test set contains 22 GM12878 cells. The number of false positive \glspl{deg} is used to measure the performance, where a lower count indicates greater robustness. The effect size analysis evaluates whether \glspl{deg} identified from imputed \gls{scrnaseq} data capture genes with both high and low \gls{lfc} in bulk RNA-seq data. Here, \gls{lfc} is defined as $\mathrm{LFC} = \log_2(y_\mathrm{target}/y_\mathrm{control})$, where $y_\mathrm{control}$ and $y_\mathrm{target}$ represent the gene expression values of the control and target groups, respectively. Genes in the upper and lower $10\,\%$ of the \gls{lfc} distribution from bulk RNA-seq data are defined as high- and low-\gls{lfc} genes, respectively. As in the \gls{de} enrichment analysis, the \gls{iou} is used to measure the overlap between high- or low-\gls{lfc} genes from bulk RNA-seq data and the \glspl{deg} from imputed \gls{scrnaseq} data.

\begin{figure*}[p]
    \centering
    \includegraphics[width=\textwidth,height=0.82\textheight,keepaspectratio]{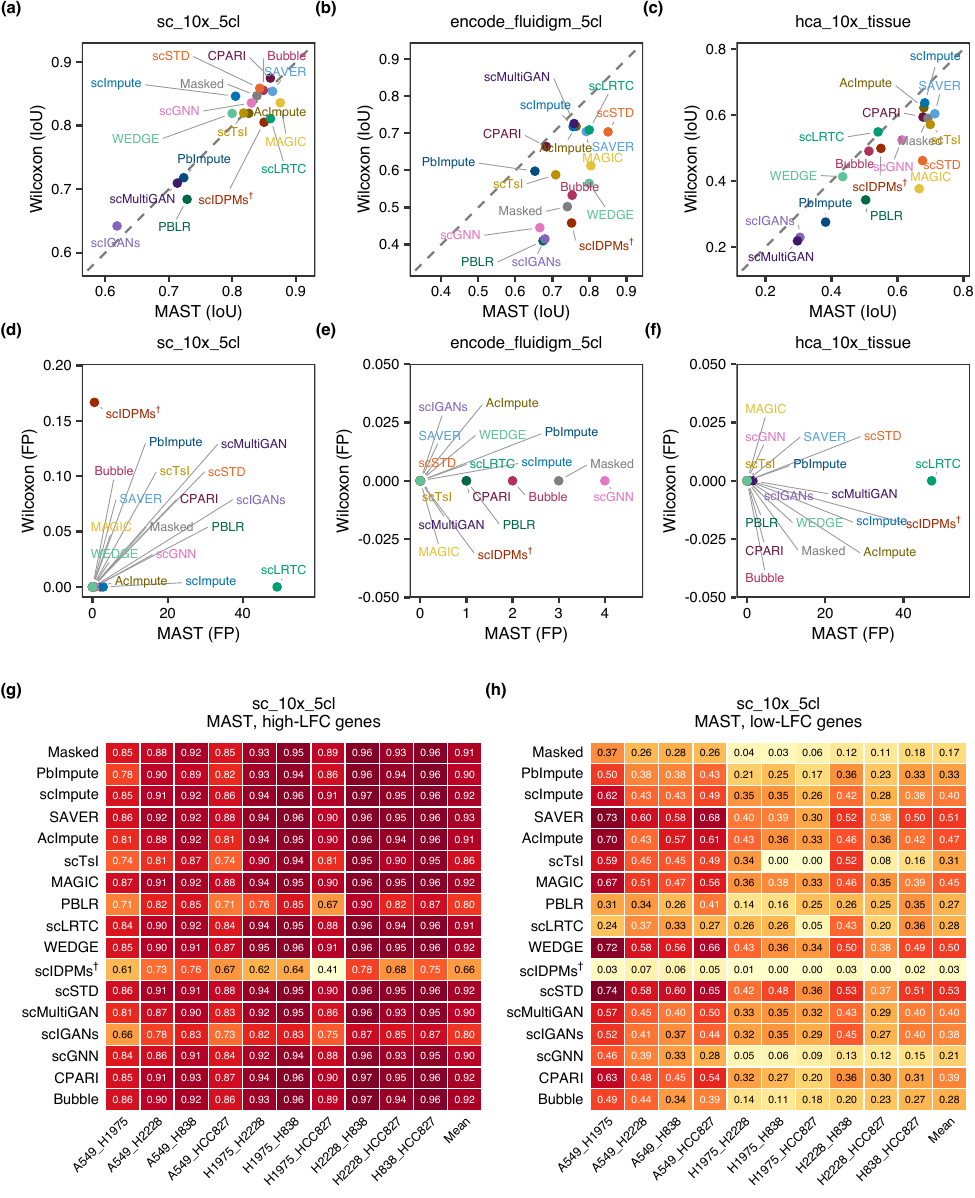}
    \caption{DE analysis performance. \textbf{(a)}--\textbf{(c)} DE enrichment analysis. IoU between DEGs identified from imputed scRNA-seq data and bulk RNA-seq data using MAST and the Wilcoxon rank-sum test. The dashed line represents equal performance. Each point represents an imputation method. \textbf{(d)}--\textbf{(f)} Null DE analysis. Average number of false positive DEGs across the tested combinations of group sizes for A549 \textbf{(d)}, GM12878 \textbf{(e)}, and monocyte \textbf{(f)}, where 6 combinations are tested for A549 and monocyte and 1 combination for GM12878. \textbf{(g)}--\textbf{(h)} Effect size analysis. IoU between DEGs identified from imputed scRNA-seq data and high- \textbf{(g)} or low- \textbf{(h)} LFC genes from bulk RNA-seq data. $^\dagger$Run with reduced settings (see \cref{subsec:method_experimental_setup}).}
    \label{fig:de_analysis_multiplot}
\end{figure*}

\namecrefs{fig:de_analysis_multiplot}~\labelcref{fig:de_analysis_multiplot}(a)--(c) show the \gls{de} enrichment analysis performance for 16 imputation methods in terms of 3 datasets, namely sc\_10x\_5cl, encode\_fluidigm\_5cl, and hca\_10x\_tissue. These datasets are selected on the basis of the availability of corresponding bulk RNA-seq data. MAST~\cite{finak2015mast} and the Wilcoxon rank-sum test~\cite{mann1947test,wilcoxon1945individual,wolf2018scanpy} are used to identify \glspl{deg} from the imputed data, and limma~\cite{ritchie2015limma} is used to identify \glspl{deg} from the bulk RNA-seq data.

The comparison of the \gls{de} enrichment analysis performance across the 16 methods evaluated on the 3 datasets shows that scImpute, SAVER, and CPARI achieve the best overall performance. In addition, 10 methods, namely PbImpute, AcImpute, scTsI, MAGIC, scLRTC, WEDGE, scIDPMs$^\dagger$, scSTD, scGNN, and Bubble, exhibit moderate performance across the 3 datasets. Conversely, PBLR, scMultiGAN, and scIGANs show the worst overall performance.

\namecrefs{fig:de_analysis_multiplot}~\labelcref{fig:de_analysis_multiplot}(d)--(f) show the null \gls{de} analysis performance using MAST~\cite{finak2015mast} and the Wilcoxon rank-sum test~\cite{mann1947test,wilcoxon1945individual,wolf2018scanpy} for 16 imputation methods in terms of 3 datasets. The analysis of false positive \glspl{deg} in the null \gls{de} analysis shows that 6 methods, namely PbImpute, SAVER, MAGIC, WEDGE, scSTD, and scTsI, produce no false positive \glspl{deg} across all datasets. In addition, 9 methods, namely scImpute, AcImpute, PBLR, scIDPMs$^\dagger$, scMultiGAN, scIGANs, scGNN, CPARI, and Bubble, produce false positive \glspl{deg} in 1 or 2 datasets. In contrast, scLRTC produces substantial false positive \glspl{deg} under MAST in 2 datasets, sc\_10x\_5cl and hca\_10x\_tissue.

\namecrefs{fig:de_analysis_multiplot}~\labelcref{fig:de_analysis_multiplot}(g) and (h) show the effect size analysis performance of high- and low-\gls{lfc} genes, respectively, for 16 imputation methods using MAST~\cite{finak2015mast} in sc\_10x\_5cl. MAST and sc\_10x\_5cl are selected as representative examples. A thorough evaluation of effect size analysis performance shows that SAVER, WEDGE, and scSTD achieve the best performance, with high \gls{iou} scores in both high- and low-\gls{lfc} genes. In addition, 12 methods, namely PbImpute, scImpute, AcImpute, scTsI, MAGIC, PBLR, scLRTC, scMultiGAN, scIGANs, scGNN, CPARI, and Bubble, exhibit moderate performance. On the other hand, scIDPMs$^\dagger$ shows the worst performance, with low \gls{iou} scores in high- or low-\gls{lfc} genes.

In summary, SAVER achieves the best overall performance, as supported by the best \gls{de} enrichment analysis performance across 3 datasets and the best effect size analysis performance, while producing no false positive \glspl{deg} in all datasets. MAGIC, WEDGE, scSTD, PbImpute, and scTsI also produce no false positive \glspl{deg} across all datasets. Conversely, PBLR, scMultiGAN, and scIGANs show the worst \gls{de} enrichment analysis performance, scLRTC produces substantial false positive \glspl{deg} under MAST, and scIDPMs$^\dagger$ shows the worst effect size analysis performance.

\subsection{Marker Gene Analysis}
\label{subsec:results_marker_genes}

\begin{figure*}[p]
    \centering
    \includegraphics[width=\textwidth,height=0.92\textheight,keepaspectratio]{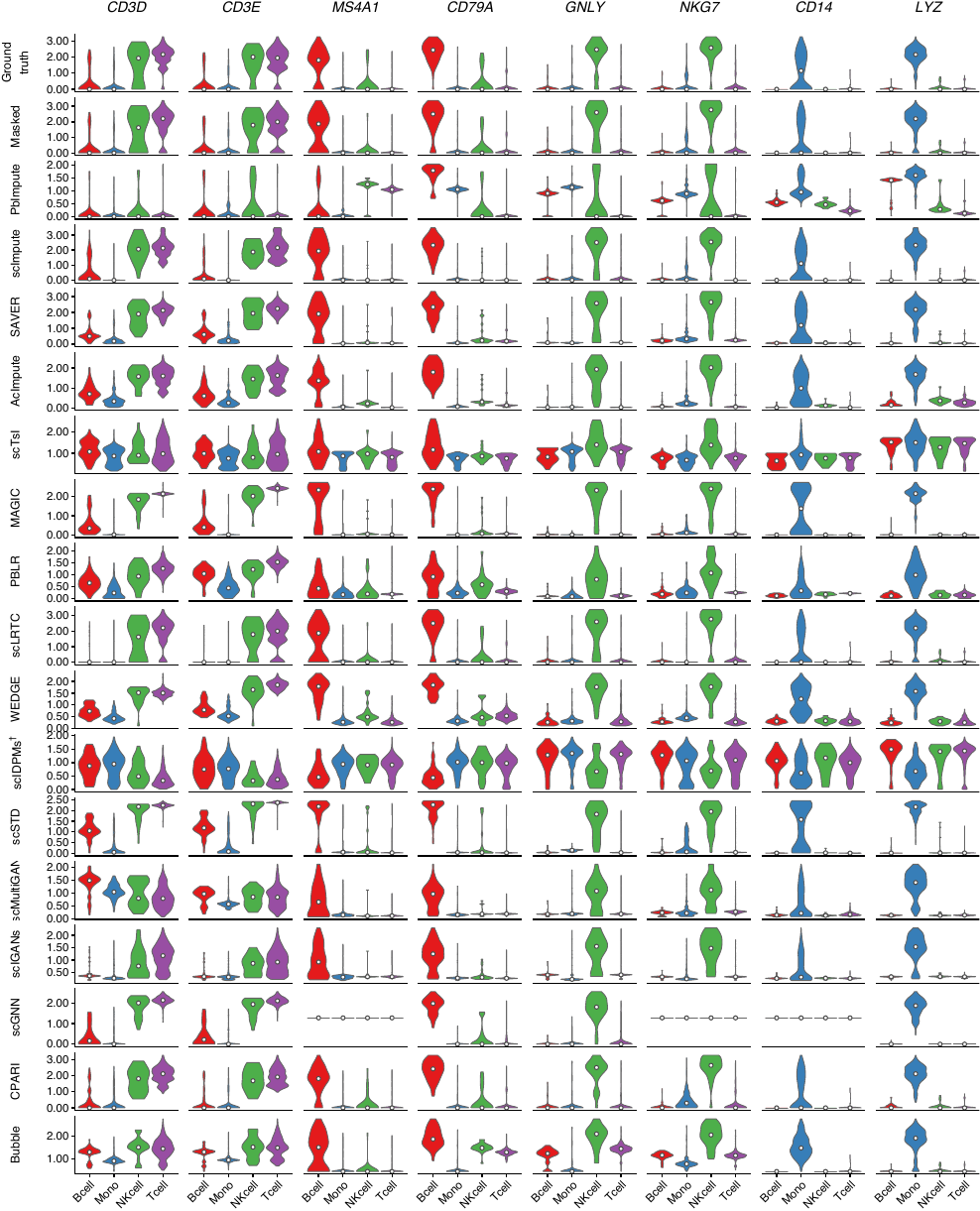}
    \caption{Comparison of imputation methods for marker gene expression in hca\_10x\_tissue. Violin plots show the distribution of expression levels for 8 marker genes across 4 cell types, T cell, B cell, NK cell, and monocyte (shown as Mono). The y-axis represents gene expression values. $^\dagger$Run with reduced settings (see \cref{subsec:method_experimental_setup}).}
    \label{fig:marker_gene_violin}
\end{figure*}

\begin{figure*}[p]
    \centering
    \includegraphics[width=\textwidth,height=0.92\textheight,keepaspectratio]{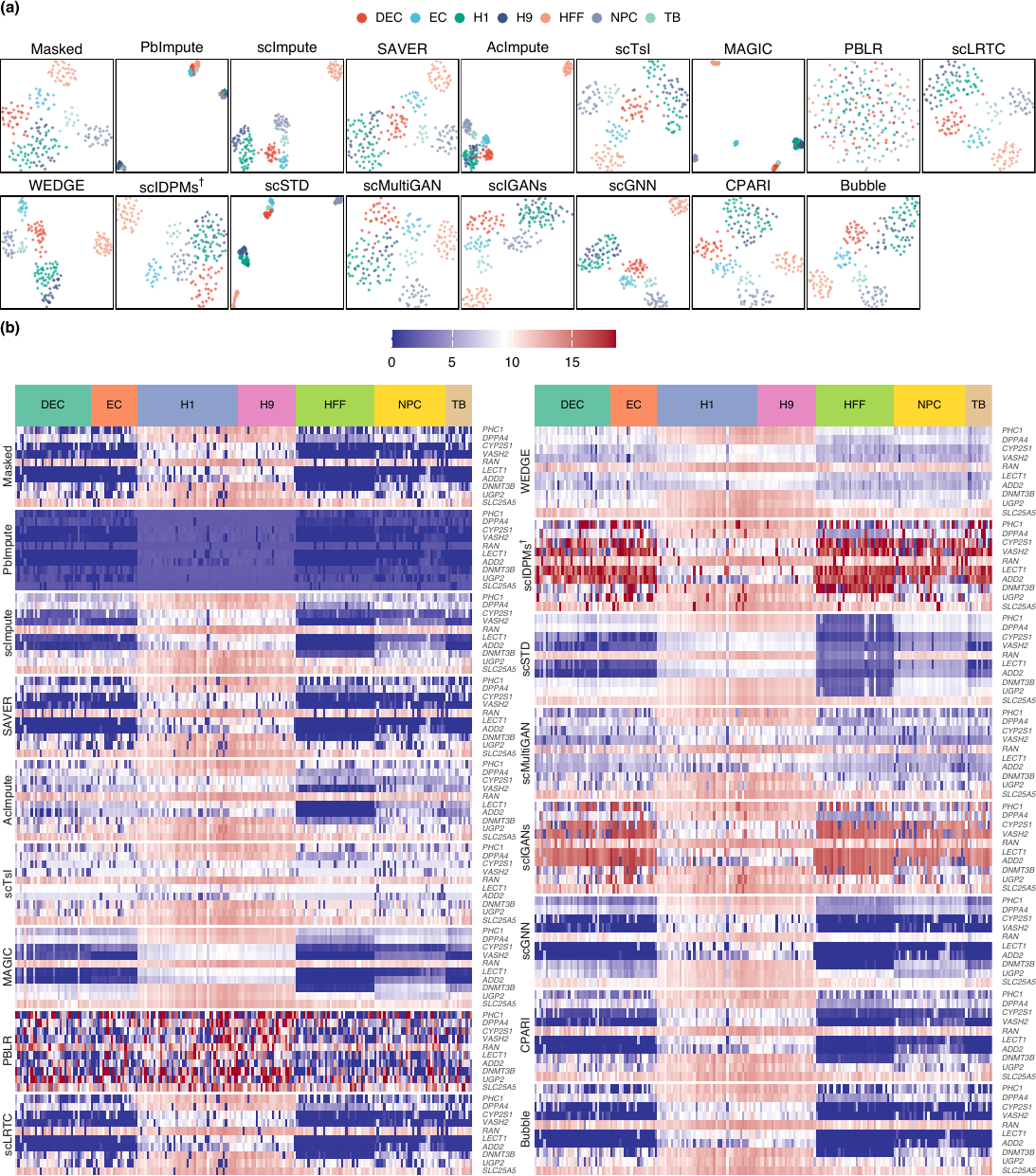}
    \caption{Marker gene expression performance on chu\_cell\_type. \textbf{(a)} UMAP visualizations with 7 cell type labels, namely DEC, EC, H1, H9, HFF, NPC, and TB, colored by cell type. Each plot represents an imputation method. \textbf{(b)} Heatmaps of 10 marker gene expression values across the 7 cell types. The x-axis represents individual cells ordered by cell type, and the y-axis represents marker genes. $^\dagger$Run with reduced settings (see \cref{subsec:method_experimental_setup}).}
    \label{fig:marker_gene_multiplot}
\end{figure*}

\cref{fig:marker_gene_violin} shows marker gene expression of 4 cell types, namely T cell, B cell, natural killer (NK) cell, and monocyte, in hca\_10x\_tissue for 16 distinct methods. Generally, \textit{CD3D} and \textit{CD3E} are considered to be marker genes for T cells, \textit{CD79A} and \textit{MS4A1} for B cells, \textit{NKG7} and \textit{GNLY} for NK cells, and \textit{CD14} and \textit{LYZ} for monocytes~\cite{cheng2023evaluating,shaffer2001signatures}. The analysis of marker gene expression reveals that scImpute, SAVER, MAGIC, scLRTC, scIGANs, and CPARI achieve the best performance, as these methods show strong expression levels of marker genes in the corresponding cell types. In addition, 5 methods, namely AcImpute, PBLR, WEDGE, scSTD, and scMultiGAN, show moderate performance, as these methods show relatively strong expression levels of marker genes for B cells and monocytes, while the marker genes for T cells, \textit{CD3D} and \textit{CD3E}, are also expressed in B cells. Conversely, 5 methods, namely PbImpute, scTsI, scIDPMs$^\dagger$, scGNN, and Bubble, show the worst performance. scTsI, scIDPMs$^\dagger$, and Bubble produce similar expression levels of the marker genes across the 4 cell types, PbImpute expresses \textit{MS4A1} more highly in NK cells and T cells than in B cells, and scGNN shows constant expression levels of 3 marker genes, namely \textit{MS4A1}, \textit{NKG7}, and \textit{CD14}, across all cells.

\namecrefs{fig:marker_gene_multiplot}~\labelcref{fig:marker_gene_multiplot}(a) and (b) show \gls{umap} visualizations with cell type labels, and marker gene expression for the 16 imputation methods in chu\_cell\_type, respectively. The \gls{umap} visualizations show that 9 methods, namely scImpute, SAVER, scTsI, scLRTC, WEDGE, scMultiGAN, scGNN, CPARI, and Bubble, separate 5 cell types, namely DEC, EC, HFF, NPC, and TB, and a group of H1 and H9, similar to the masked data. In addition, 4 methods, namely MAGIC, scIDPMs$^\dagger$, scSTD, and scIGANs, show moderate separation. MAGIC and scSTD separate the cell types but compress the cells of each cell type into small clusters, and scIDPMs$^\dagger$ and scIGANs do not separate NPC from the group of H1 and H9. Conversely, 3 methods, namely PbImpute, AcImpute, and PBLR, do not show clear separation. PbImpute merges DEC, EC, HFF, and TB into 1 cluster and H1, H9, and NPC into another, AcImpute merges the 6 cell types other than HFF into 1 cluster, and PBLR mixes the 7 cell types together. The analysis of the marker gene expression shows that 8 methods, namely scImpute, SAVER, MAGIC, scLRTC, scSTD, scGNN, CPARI, and Bubble, exhibit distinct marker gene expression patterns similar to the masked data, with low expression levels of 4 marker genes, namely \textit{CYP2S1}, \textit{VASH2}, \textit{LECT1}, and \textit{ADD2}, in the 5 cell types other than H1 and H9. In addition, 4 methods, namely AcImpute, WEDGE, scMultiGAN, and scTsI, show moderate marker gene expression patterns. AcImpute retains the low expression levels of the 4 marker genes only in HFF, WEDGE and scMultiGAN show smaller differences between H1 and H9 and the other cell types, and scTsI shows almost no differences across the 7 cell types. In contrast, 4 methods, namely PbImpute, PBLR, scIDPMs$^\dagger$, and scIGANs, show the worst performance. PbImpute shows uniformly low expression levels across the 7 cell types, PBLR shows no cell-type-specific patterns, and scIDPMs$^\dagger$ and scIGANs show higher expression levels of the 4 marker genes in DEC, EC, and HFF than in H1 and H9.

In summary, the marker gene analysis reveals substantial variability in the ability of imputation methods to preserve biologically meaningful gene expression patterns across the 2 datasets. scImpute, SAVER, scLRTC, and CPARI show the best overall performance, as they consistently exhibit cell-type-specific marker gene expression in hca\_10x\_tissue and distinct marker gene expression patterns in chu\_cell\_type, which is further supported by clear cell type separation in the \gls{umap} visualizations. Conversely, PbImpute shows the worst results in both datasets, followed by scIDPMs$^\dagger$ and PBLR, which show the worst performance in 2 of the 3 analyses.

\subsection{Trajectory Analysis}
\label{subsec:results_trajectory_analysis}

\begin{figure*}[p]
    \centering
    \includegraphics[width=\textwidth,height=0.92\textheight,keepaspectratio]{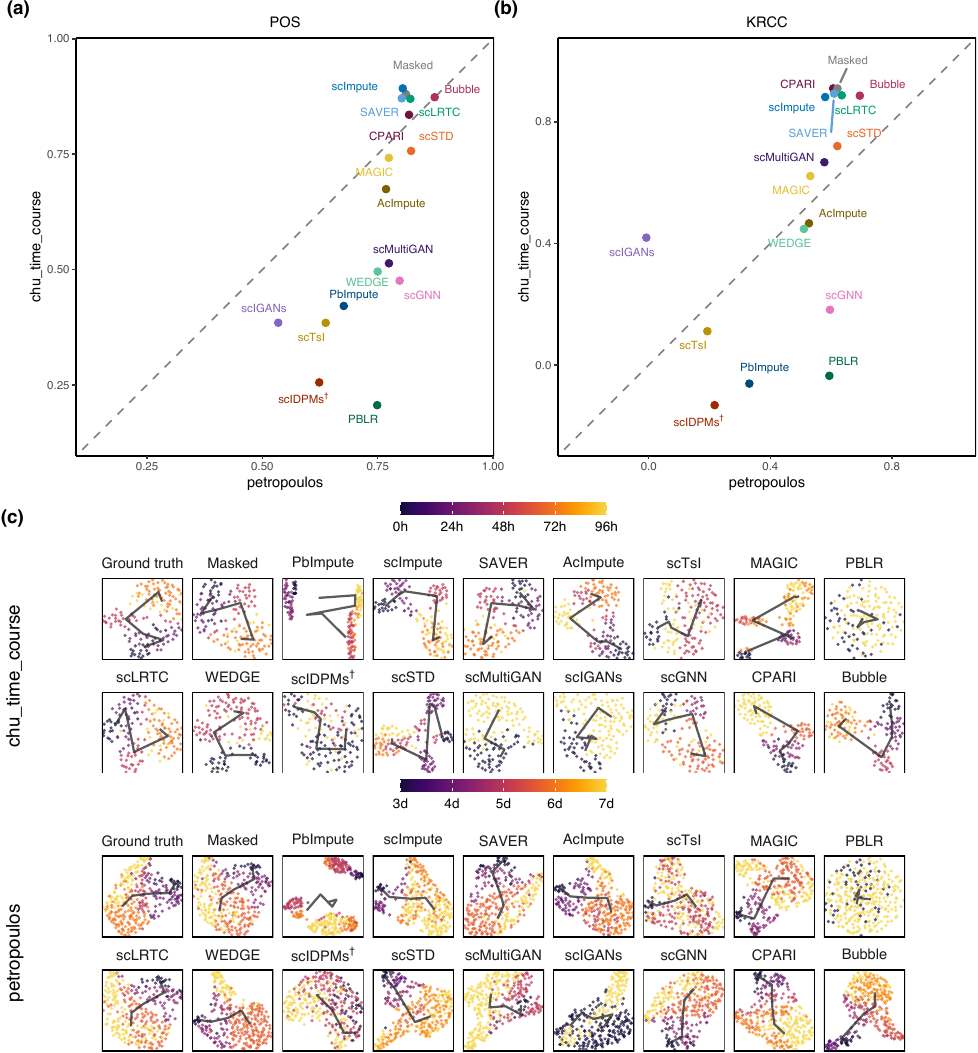}
    \caption{Trajectory analysis performance. \textbf{(a)}--\textbf{(b)} POS and KRCC between the inferred pseudotime from imputed data and the true cellular development time label, respectively. The x-axis and y-axis represent the performance of petropoulos and chu\_time\_course, respectively. The dashed line represents equal performance across the 2 datasets. Each point in the plot represents an imputation method. \textbf{(c)} UMAP visualizations with inferred pseudotime trajectories for chu\_time\_course (top) and petropoulos (bottom). Cells are colored by inferred pseudotime. $^\dagger$Run with reduced settings (see \cref{subsec:method_experimental_setup}).}
    \label{fig:trajectory_analysis_multiplot}
\end{figure*}

\namecrefs{fig:trajectory_analysis_multiplot}~\labelcref{fig:trajectory_analysis_multiplot}(a) and (b) show \gls{pos} and \gls{krcc} for 16 imputation methods in terms of 2 datasets, namely petropoulos and chu\_time\_course. High \gls{pos} and \gls{krcc} represent proximity to true cell development labels. The comparison of \gls{pos} and \gls{krcc} across 16 methods evaluated on 2 datasets shows that 5 methods, namely scImpute, SAVER, scLRTC, CPARI, and Bubble, achieve relatively high performance in both datasets. In addition, 6 methods, namely AcImpute, MAGIC, WEDGE, scSTD, scMultiGAN, and scGNN, show moderate performance. Conversely, 5 methods, namely PbImpute, scTsI, PBLR, scIDPMs$^\dagger$, and scIGANs, show the worst performance. However, the masked baseline shows relatively high performance, and none of the 16 methods exceed it in both datasets in terms of either \gls{pos} or \gls{krcc}.

\cref{fig:trajectory_analysis_multiplot}(c) shows \gls{umap} visualizations of trajectory analysis for 16 imputation methods in terms of 2 datasets. The qualitative analyses through the \gls{umap} visualizations show that the pseudotime changes gradually for scImpute, SAVER, scLRTC, CPARI, and Bubble, with cells colored by pseudotime progressing smoothly. On the other hand, the \gls{umap} visualizations of PBLR, scIDPMs$^\dagger$, and scIGANs in both datasets, and those of scTsI and scMultiGAN in chu\_time\_course, show that the pseudotime does not change smoothly, with cells of different time points appearing intermixed. In addition, PbImpute divides the cells into separate clusters in both datasets. However, since trajectory analysis is performed in higher dimensions than \gls{umap}, \gls{pos} and \gls{krcc} results may not be fully reflected in the 2-dimensional \gls{umap} projections.

In summary, the trajectory analysis reveals that 5 methods, namely scImpute, SAVER, scLRTC, CPARI, and Bubble, consistently preserve the temporal ordering of cells across 2 datasets, while PbImpute, scTsI, PBLR, scIDPMs$^\dagger$, and scIGANs perform the worst. However, the fact that none of the 16 methods outperforms the masked baseline in both datasets highlights a critical challenge that imputation can distort the underlying developmental structure of \gls{scrnaseq} data. This can potentially lead to less accurate trajectory analysis than simply using the data without imputation.

\subsection{Cell Type Annotation}
\label{subsec:results_cell_type_annotation}

\begin{figure*}[p]
    \centering
    \includegraphics[width=\textwidth,height=0.86\textheight,keepaspectratio]{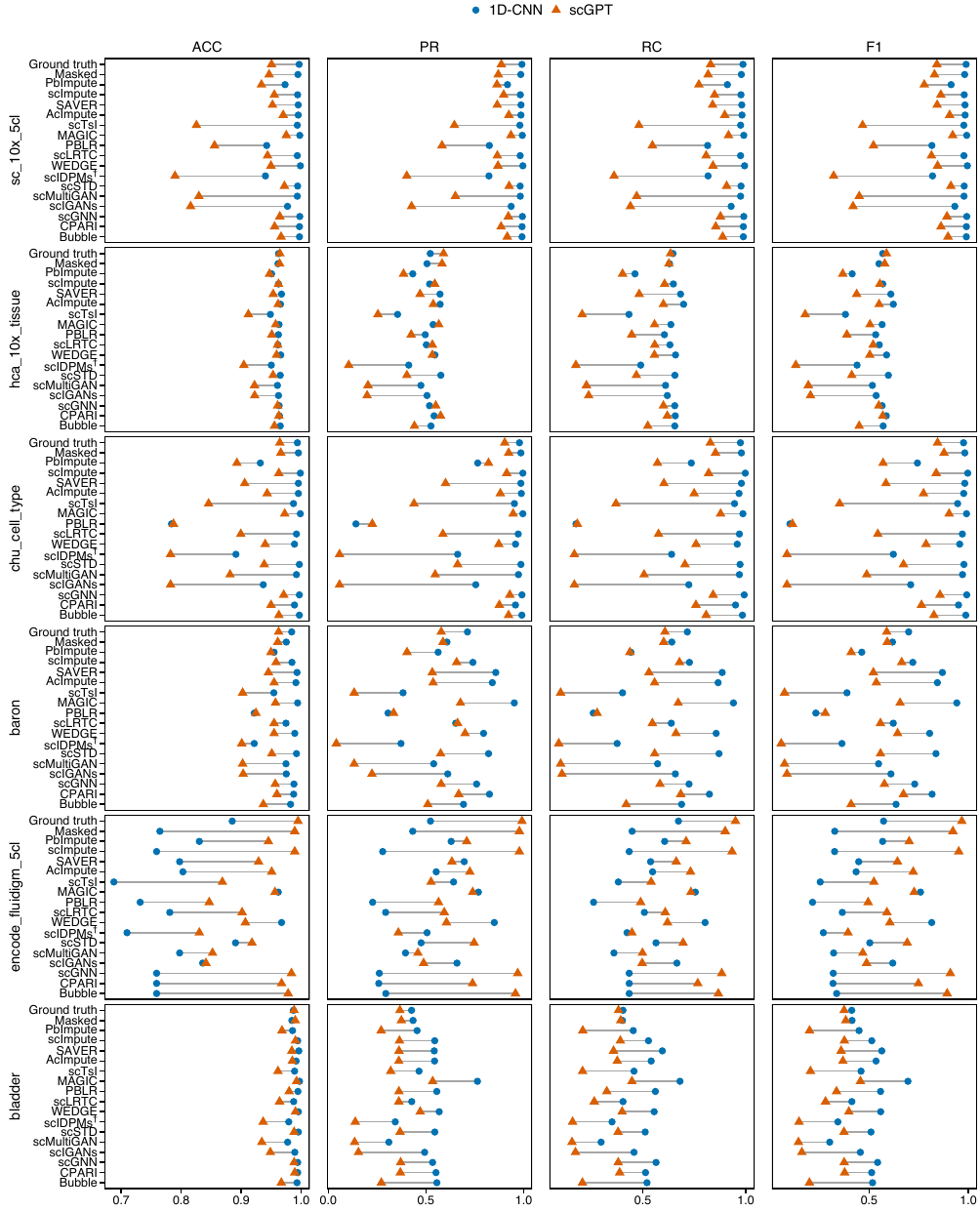}
    \caption{Cell type annotation performance. Dumbbell plots comparing ACC, PR, RC, and F1 achieved by 1D-CNN (blue circles) and scGPT~\cite{cui2024scgpt} (orange triangles) across 6 datasets, including sc\_10x\_5cl, hca\_10x\_tissue, chu\_cell\_type, baron, encode\_fluidigm\_5cl, and bladder. The x-axis represents the value of each evaluation measure, and the y-axis represents different imputation methods. Each horizontal line connects the 1D-CNN and scGPT performances for a given method. $^\dagger$Run with reduced settings (see \cref{subsec:method_experimental_setup}).}
    \label{fig:cell_type_annotation_multiplot}
\end{figure*}

\cref{fig:cell_type_annotation_multiplot} shows the \gls{acc}, \gls{pr}, \gls{rc}, and \gls{f1} of cell type annotation for the 16 imputation methods in terms of 6 datasets, namely sc\_10x\_5cl, hca\_10x\_tissue, chu\_cell\_type, baron, encode\_fluidigm\_5cl, and bladder, using \gls{1d-cnn} and scGPT~\cite{cui2024scgpt} for annotation. Higher \gls{acc} represents better overall accuracy of cell type annotations, higher \gls{pr} represents better precision in identifying true cell types, higher \gls{rc} represents better recall of true cell types, and higher \gls{f1} represents a better balance between \gls{pr} and \gls{rc}.

A thorough analysis of \gls{acc}, \gls{pr}, \gls{rc}, and \gls{f1} for the 16 imputation methods in terms of 6 datasets reveals that MAGIC achieves the best overall performance. In addition, 14 methods, namely PbImpute, scImpute, SAVER, AcImpute, scTsI, PBLR, scLRTC, WEDGE, scSTD, scMultiGAN, scIGANs, scGNN, CPARI, and Bubble, show moderate results. On the other hand, scIDPMs$^\dagger$ shows the worst performance. The comparison of the 2 cell type annotation methods, namely \gls{1d-cnn} and scGPT, shows that \gls{1d-cnn} outperforms scGPT in 5 datasets, namely sc\_10x\_5cl, hca\_10x\_tissue, chu\_cell\_type, baron, and bladder.

In summary, MAGIC shows the best cell type annotation performance, while scIDPMs$^\dagger$ shows the worst performance. Furthermore, the choice of cell type annotation methods introduces considerable variability, as \gls{1d-cnn} outperforms scGPT in 5 of the 6 datasets.

\section{Discussion}
\label{sec:discussion}

\input{tables/performance_summary}

In this study, we conduct a systematic evaluation of 16 imputation methods across 6 downstream tasks, including numerical gene expression recovery, cell clustering, \gls{de} analysis, marker gene analysis, trajectory analysis, and cell type annotation. The evaluation is performed using 26 real and 4 simulated datasets generated from 10 different protocols, including 10x Chromium, CEL-seq2, SMART-seq2, SMART-seq, Drop-seq, STRT-seq, inDrop, Fluidigm C1, Microwell-seq, and Sort-seq. The results reveal substantial variability in the performance of imputation methods across different downstream tasks and datasets, which highlights the importance of carefully selecting imputation methods based on specific downstream analyses and dataset characteristics.

\cref{tab:performance_summary} summarizes the performance of the 16 imputation methods across 6 downstream tasks. SAVER shows the best overall performance with the highest performance in 3 tasks, namely \gls{de} analysis, marker gene analysis, and trajectory analysis, and moderate performance in the remaining 3 tasks. In addition, scImpute, MAGIC, scLRTC, WEDGE, CPARI, and Bubble show moderately high overall performance with the best performance in 1 to 3 tasks, and the worst performance in at most 1 task. Similarly, PbImpute, AcImpute, scTsI, scSTD, scMultiGAN, and scGNN show moderately low overall performance, with no best performance in any task, and the worst performance in 0 to 2 tasks. Conversely, PBLR, scIDPMs$^\dagger$, and scIGANs show the worst overall performance, with the worst performance in 3 to 6 tasks. We find that 5 of the 9 traditional methods, namely SAVER, scImpute, MAGIC, scLRTC, and WEDGE, show the best or moderately high overall performance, whereas only 2 of the 7 \gls{dl}-based methods, namely CPARI and Bubble, reach the same level of overall performance. This suggests that traditional methods may be more effective at preserving biologically meaningful information across a wide range of downstream analyses, while the performance of \gls{dl}-based methods can be more variable and may require careful tuning and validation for specific tasks.

In numerical gene expression recovery, the performance of imputation methods varies substantially across methods. WEDGE achieves the best overall \gls{lnd} performance, with medians closest to 0 in the largest number of datasets, followed by PBLR. WEDGE shows superior \gls{lnd} performance likely due to its biased low-rank matrix-based approach that assigns low weights to 0 elements and minimizes approximation error for nonzero elements. This approach effectively separates true biological signal from dropouts without over-imputing zero entries. WEDGE also exhibits the lowest \gls{mae} and \gls{medae}, which further supports this finding. PBLR preserves the data structure through its cell sub-population-based bounded low-rank matrix recovery. However, the optimization of PBLR does not converge within the default maximum of 200 iterations in some cell subpopulations, which yields extremely large imputed values in 6 datasets, namely hcc, chu\_cell\_type, petropoulos, chu\_time\_course, usoskin, and encode\_fluidigm\_5cl. Protocol-wise analysis reveals that WEDGE maintains \gls{lnd} values closest to zero only in CEL-seq2 and Drop-seq, which indicates that its performance depends on the protocol. For comparison with bulk RNA-seq data, WEDGE also achieves the best overall performance, and AcImpute and MAGIC also show the highest correlation at the pseudo-bulk and cell line levels, respectively, through their diffusion-based smoothing, which enforces locally coherent expression profiles aligned with averaged bulk patterns. In contrast, scIGANs exhibits the worst correlation with bulk RNA-seq data at both the pseudo-bulk and cell line levels despite its moderate numerical recovery performance from ground truth data. This highlights a trade-off between numerical accuracy and biological fidelity. Its image-based architecture, which reshapes the expression vector of each cell into a square image, may not preserve the relative expression levels across genes that determine the correlation with bulk RNA-seq data. The over-imputation by scIDPMs$^\dagger$ may arise from its iterative denoising process that pushes sparse dropout entries toward higher-density non-zero modes across multiple sequential denoising steps, which also results in the highest \gls{mae} and \gls{medae}, although its reduced settings may also contribute to this result. scLRTC consistently under-imputes because its low-rank matrix-based approach compresses the dynamic range of highly variable genes. Furthermore, methods show largely consistent behavior on 10x Chromium, CEL-seq2, and Drop-seq datasets, while SMART-seq, SMART-seq2, and Fluidigm C1 datasets demonstrate higher instability. This is likely because the latter 3 protocols use only read-counts without \glspl{umi}. The absence of \glspl{umi} leads to increased technical noise that makes accurate imputation more challenging~\cite{cheng2023evaluating,ziegenhain2017comparative}.

In cell clustering, the performance of imputation methods also varies substantially across methods. In terms of consistency, scLRTC achieves the best performance. This is likely because scLRTC reconstructs gene expression values through its low-rank tensor completion that captures global gene expression patterns while preserving the distinct expression signatures that define cell clusters. In terms of coherency, MAGIC achieves the best performance. MAGIC produces tightly grouped clusters, likely because its diffusion-based smoothing over \gls{knn} graphs harmonizes expression profiles within cell neighborhoods, which enhances intra-cluster homogeneity. The distinction between consistency and coherency suggests that well-separated clusters do not necessarily correspond to accurately recovered expression patterns, as MAGIC achieves the best cluster coherency but not the best cluster consistency. Conversely, PBLR shows the worst results in both consistency and coherency, and scIDPMs$^\dagger$ shows the worst coherency. For PBLR, this may be due to its bounded low-rank matrix recovery approach that compresses expression variability into a small number of latent factors and can merge expression signatures of distinct but transcriptionally similar cell populations, as well as the extremely large imputed values caused by its non-converged optimization. For scIDPMs$^\dagger$, its over-imputation may obscure expression differences between closely related cell populations, which leads to poor cluster separation, although its reduced settings may also contribute to this result. Notably, none of the 16 methods exceed the \gls{ari} scores of the masked baseline in 3 datasets, and at least half of the 16 methods do not exceed it in 25 of the 30 datasets, which suggests that imputation does not universally improve cell clustering and can even degrade cluster consistency by introducing imputed expression patterns. Furthermore, the \gls{umap} visualization reveals that WEDGE maintains visually distinct clusters across 3 of the 4 simulated datasets with varying dropout rates, while the remaining 15 methods show limited ability to recover cluster structures. This suggests that these 15 methods may be less robust to dropout-induced sparsity, which can lead to poor cluster recovery in datasets with high dropout rates.

In \gls{de} analysis, SAVER achieves the best overall performance, with the best \gls{de} enrichment analysis performance across 3 datasets and the best effect size analysis performance. This is likely due to its model-based approach that estimates the true expression level of each gene by borrowing information from other genes, which preserves the relative expression differences between cell populations. SAVER also produces no false positive \glspl{deg} in all datasets, which suggests that its posterior mean estimates recover expression values without introducing systematic expression patterns between randomly partitioned cell groups. In contrast, scLRTC produces substantial false positive \glspl{deg} under MAST~\cite{finak2015mast} in 2 datasets, sc\_10x\_5cl and hca\_10x\_tissue. This indicates that its low-rank tensor completion approach can introduce imputed expression patterns that inflate false positive rates. PbImpute, MAGIC, WEDGE, scSTD, and scTsI also produce no false positive \glspl{deg} across all datasets. However, PbImpute, MAGIC, and scTsI show moderate \gls{de} enrichment and effect size analysis performance. The low false positive rates of these methods do not universally translate into accurate \gls{deg} identification, which indicates that avoiding false positives is necessary but not sufficient for accurate \gls{de} analysis. Conversely, PBLR, scMultiGAN, and scIGANs show the worst \gls{de} enrichment analysis performance. For PBLR, this indicates that its bounded low-rank matrix recovery fails to preserve the relative expression differences between cell populations, which may be further affected by its non-converged optimization. For the \gls{gan}-based methods, scMultiGAN and scIGANs, the generation of imputed values from learned data distributions may not preserve the relative expression differences between cell populations. In addition, scIDPMs$^\dagger$ shows the worst effect size analysis performance, likely because its over-imputation compresses the expression differences of genes with high or low \gls{lfc}.

In marker gene analysis, we examine the preservation of biologically meaningful gene expression patterns by comparing marker gene expression levels across different cell types in imputed data. The results reveal that scImpute, SAVER, scLRTC, and CPARI consistently show the best performance, with strong cell-type-specific marker gene expression patterns across the 2 datasets. scImpute shows the best performance, likely due to its mixture distribution-based approach that selectively imputes values on a per-gene basis, which allows it to impute only the values that are likely to be technical dropouts while preserving biological zeros. This approach maintains the distinct marker gene expression patterns in their respective cell types while keeping non-expressing cell types at low expression levels. SAVER similarly estimates the true expression level of each gene on a per-gene basis by borrowing information from other genes, which retains the relative expression differences of marker genes between cell types. scLRTC preserves cell-type-specific expression signatures through its low-rank tensor completion that uses cell-cell similarity information, and CPARI imputes dropout events within each cell cluster, which may prevent marker gene expression from spreading across cell types. Conversely, PbImpute shows the worst results across both datasets, followed by scIDPMs$^\dagger$ and PBLR, as they produce indistinct marker gene expression patterns and poor cell type separation. For PbImpute, its static repair step that adjusts outlying nonzero values may suppress the distinct expression peaks of marker genes. For scIDPMs$^\dagger$, its over-imputation can raise the expression of marker genes in non-expressing cell types, which blurs cell-type-specific signatures. For PBLR, its bounded low-rank matrix recovery approach compresses gene expression variability into a small number of latent factors, which in turn weakens the distinct expression peaks of marker genes and blurs cell-type-specific signatures. Notably, scGNN and scIGANs perform well in 1 of the 2 datasets but poorly in the other, while scLRTC under-imputes in numerical gene expression recovery yet shows the best marker gene analysis performance. This inconsistency between numerical gene expression recovery and marker gene analysis further reinforces the finding that overall imputation accuracy does not guarantee the preservation of biologically meaningful expression patterns.

In trajectory analysis, we evaluate the ability of imputation methods to preserve the temporal ordering of cells by comparing inferred pseudotime from imputed data with true cellular development time labels. The results reveal that scImpute, SAVER, scLRTC, CPARI, and Bubble consistently preserve the temporal ordering of cells across the 2 datasets. This shows that 2 model-based methods, 1 low-rank matrix-based method, and 2 \gls{ae}-based methods perform well in trajectory analysis. This may be due to the fact that these methods preserve the relative expression differences along developmental trajectories, which is critical for accurate pseudotime inference. The model-based methods, scImpute and SAVER, estimate expression values on a per-gene basis using statistical models of the observed counts, which helps maintain the gradual expression changes that define developmental trajectories. scLRTC reconstructs gene expression values through its low-rank tensor completion that captures global gene expression patterns while preserving the distinct expression signatures that define developmental trajectories. The \gls{ae}-based methods, CPARI and Bubble, learn latent representations using \gls{ae} architectures that retain the continuous structure of developmental progression without collapsing intermediate states. Conversely, PbImpute, scTsI, PBLR, scIDPMs$^\dagger$, and scIGANs show the worst performance. PbImpute divides the cells into separate clusters, which may disrupt the continuous pseudotime ordering. For scIDPMs$^\dagger$, its iterative denoising process trained to denoise expression values toward high-density modes may collapse the subtle expression gradients between adjacent developmental stages into discrete expression states, although its reduced settings may also contribute to this result. For scIGANs, its adversarial training aims to match the overall data distribution, but the \gls{gan}-based generation process may suffer from mode collapse that concentrates imputed values around high-density expression states rather than preserving the continuous gradients between developmental stages. This is consistent with the UMAP visualizations, where cells of different time points appear intermixed for PBLR, scIDPMs$^\dagger$, and scIGANs in both datasets and for scTsI in chu\_time\_course. Furthermore, none of the 16 methods outperform the masked baseline in both datasets, which highlights that imputation can distort the underlying developmental structure of \gls{scrnaseq} data and lead to less accurate trajectory inference than simply using the original data. This suggests that trajectory analysis, which depends on the preservation of continuous expression gradients rather than discrete cluster boundaries, is particularly sensitive to imputation artifacts that alter the relative ordering of expression values along developmental axes.

In cell type annotation, we assess the impact of imputation on the accuracy of cell type predictions by comparing annotations derived from imputed data with known cell type labels. The results reveal that MAGIC achieves the best overall performance across the 6 datasets. The strong performance of MAGIC may be attributed to its smoothing-based approach that propagates expression signals across similar cells, which enhances the expression profiles used by annotation classifiers to distinguish cell types without the elimination of cell-type-specific patterns. Conversely, scIDPMs$^\dagger$ shows the worst performance. This may be due to its iterative denoising process that over-imputes expression values toward high-density modes, which can blur the boundaries between transcriptionally similar cell types, although its reduced settings may also contribute to this result. The comparison between \gls{1d-cnn} and scGPT reveals that \gls{1d-cnn} outperforms scGPT in 5 of the 6 datasets. This suggests that the choice of cell type annotation methods introduces considerable variability, and that scGPT, the only foundation model evaluated in this study, does not necessarily outperform a simple classifier trained on each dataset.

\section{Conclusions}
\label{sec:conclusions}

Overall, the comprehensive evaluation across 6 downstream tasks reveals that although some imputation methods excel in specific tasks, there is no universally superior method across all tasks. This indicates that the choice of imputation method should be carefully tailored to the specific downstream task and dataset characteristics to ensure optimal performance. For instance, SAVER and scImpute are recommended for tasks that require the preservation of biologically meaningful information, such as \gls{de} analysis, marker gene analysis, and trajectory analysis, MAGIC for cell type annotation, while WEDGE may be more suitable for tasks that prioritize numerical gene expression recovery. The variability in performance across different tasks and datasets also suggests that researchers should consider using multiple imputation methods and comparing their results to ensure the robustness of their findings. In addition, while traditional methods, such as SAVER, scImpute, and MAGIC, show relatively better performance across a wide range of tasks, the performance of \gls{dl}-based methods is more variable, with 2 methods, namely CPARI and Bubble, showing moderately high performance, and 2 methods, namely scIDPMs$^\dagger$ and scIGANs, showing the worst performance across the 6 tasks. This suggests that further development and optimization of \gls{dl}-based imputation methods are needed to achieve consistent performance across diverse downstream analyses. The trade-off between numerical gene expression recovery performance and biological signal preservation is a critical consideration in the design and selection of imputation methods, as methods that excel in one aspect may perform poorly in the other, which can have important implications for the interpretation of \gls{scrnaseq} data and the biological conclusions drawn from it.

Moreover, this study includes limitations that should be acknowledged. One limitation is the selection of imputation methods, which covers a range of recently developed approaches and 3 well-known traditional methods, i.e., scImpute, MAGIC, and SAVER, but does not encompass all existing methods evaluated in previous benchmarking studies. In addition, all methods are run with the default settings of their original implementations without hyperparameter tuning, and scIDPMs$^\dagger$ is run with reduced settings due to computational constraints. Therefore, the results may not reflect the best achievable performance of each method, particularly for scIDPMs$^\dagger$. Furthermore, while we evaluate performance in terms of 26 real and 4 simulated datasets with 10 different protocols, it may be beneficial to further expand the diversity of datasets, including more complex tissues and disease states. In terms of downstream tasks, we focus on 6 core tasks, including numerical gene expression recovery, cell clustering, \gls{de} analysis, marker gene analysis, trajectory analysis, and cell type annotation. Future studies could explore additional open problems, such as RNA velocity analysis and batch effect correction, to further understand the impact of imputation on more complex analyses.

\section*{Acknowledgments}

\subsection*{Author Contributions}
Y.I.: Conceptualization, Data curation, Formal analysis, Investigation, Methodology, Software, Validation, Visualization, and Writing -- original draft. A.F.A.: Conceptualization, Data curation, Methodology, Supervision, Validation, Visualization, Writing -- original draft, and Writing -- review \& editing. K.K.: Funding acquisition, Resources, and Writing -- review \& editing. A.D.: Funding acquisition, Resources, Supervision, and Writing -- review \& editing. M.N.A.: Conceptualization, Funding acquisition, Supervision, and Writing -- review \& editing.

\subsection*{Funding}
This work was supported by Japan Science and Technology Agency (JST) as part of Adopting Sustainable Partnerships for Innovative Research Ecosystem (ASPIRE), Grant Number JPMJAP2403.

\subsection*{Conflicts of Interest}
The authors declare that there is no conflict of interest regarding the publication of this article.

\subsection*{AI Tool Usage Declaration}
During the preparation of this work, the authors used Anthropic Claude to assist in implementing the analysis code and to improve the language, clarity, and presentation of the manuscript. The authors take full responsibility for the code and the final manuscript.

\subsection*{Data Availability}
All datasets used in this study are publicly available. See \cref{tab:dataset_details} for more details. The codes are available from the corresponding authors on reasonable request.

\section*{Supplementary Materials}

Fig. S1. Cell clustering \gls{umap} visualization across all datasets.

Table S1. Detailed numerical recovery performance.

Table S2. Detailed cell clustering performance.

Table S3. Detailed \gls{de} enrichment and null \gls{de} analysis performance.

Table S4. Detailed \gls{de} effect size analysis performance.

Table S5. Detailed trajectory analysis performance.

Table S6. Detailed cell type annotation performance.

Table S7. Implementation and parameter settings of the imputation methods.

\printbibliography

\end{document}

%% file: tables/existing_benchmarks.tex
\begin{table*}[tb]
\renewcommand{\arraystretch}{1.6}
\setlength{\aboverulesep}{0pt}
\setlength{\belowrulesep}{0pt}
\caption{Summary of existing benchmarking studies on scRNA-seq data imputation methods}
\label{tab:related_work}
\centering
\fontsize{7pt}{9pt}\selectfont
\setlength{\tabcolsep}{4pt}
\begin{tabular}{l ccc cccc ccc}
\toprule
\multirow{3}{*}{Study}
  & \multicolumn{7}{c}{Methods}
  & \multirow{3}{*}{Datasets}
  & \multirow{3}{*}{Protocols}
  & \multirow{3}{*}{Tasks} \\
\cmidrule(lr){2-8}
  & \multicolumn{3}{c}{Traditional}
  & \multicolumn{4}{c}{DL-based}
  & & & \\
\cmidrule(lr){2-4}\cmidrule(lr){5-8}
  & \makecell{Model-\\based}
  & \makecell{Smoothing-\\based}
  & \makecell{Low-rank\\Matrix-based}
  & \makecell{Diffusion-\\based}
  & \makecell{GAN-\\based}
  & \makecell{GNN-\\based}
  & \makecell{AE-\\based}
  & & & \\
\midrule
\citet{hou2020systematic}   & 6 & 3 & 3 & 0 & 0 & 0 & 6 & 16 & 5 & 4 \\
\citet{dai2022scimc}   & 2 & 3 & 2 & 0 & 1 & 1 & 3 & 8 & 1 & 4 \\
\citet{cheng2023evaluating} & 2 & 4 & 1 & 0 & 0 & 0 & 4 & 16 & 3 & 3 \\
\midrule
\rowcolor{gray!15}
This Study    & 3 & 3 & \textbf{3}
  & \textbf{2} & \textbf{2} & \textbf{1} & 2
  & \textbf{30} & \textbf{10} & \textbf{6} \\
\bottomrule
\end{tabular}
\end{table*}

%% file: tables/datasets.tex
\begin{table*}[p]
\caption{Details of the used scRNA-seq datasets}\label{tab:dataset_details}
\centering
\fontsize{6.5pt}{7.8pt}\selectfont
\renewcommand{\arraystretch}{1.0}
\setlength{\tabcolsep}{3pt}
\renewcommand*{\multicitedelim}{\addcomma\addnbspace}
\begin{threeparttable}
\begin{tabular*}{\textwidth}{@{\extracolsep{\fill}}>{\raggedright\arraybackslash}p{2.9cm} >{\raggedright\arraybackslash}p{4.05cm} >{\raggedright\arraybackslash}p{52.5pt} >{\raggedleft\arraybackslash}p{61pt} >{\raggedleft\arraybackslash}p{40pt} >{\raggedright\arraybackslash}p{52pt}@{}}
\toprule
\textbf{Dataset} & \textbf{Description} & \textbf{Source} & \makecell[tr]{\textbf{Size}\\\textbf{($\mathrm{\mathbf{Cells}} \times \mathrm{\mathbf{Genes}}$)}} & \makecell[tr]{\textbf{Sparsity}\\\textbf{Rates (\%)}} & \textbf{Protocol} \\
ad\_case~\cite{grubman2019singlecell} & Human brain cells with Alzheimer's disease & GSE138852 & $13214 \times 10850$ & $93.53$ & 10x Chromium \\
jurkat & Human Jurkat T-cell leukemia cell line scRNA-seq dataset & 10x Genomics\tnote{a} & $3234 \times 15235$ & $79.10$ & 10x Chromium \\
293t & Human HEK293T embryonic kidney cell line scRNA-seq dataset & 10x Genomics\tnote{b} & $2868 \times	16290$ & $79.11$ & 10x Chromium \\
pbmc4k & Peripheral blood mononuclear cells (PBMCs) from a healthy donor. & 10x Genomics\tnote{c} & $4220 \times 16412$ & $92.26$ & 10x Chromium \\
sc\_10x~\cite{tian2019benchmarking,tian2018scpipe} & Single cells from three human lung adenocarcinoma cell lines & GSM3022245 & $902 \times 16468$ & $45.02$ & 10x Chromium \\
sc\_10x\_5cl~\cite{tian2019benchmarking,tian2018scpipe} & Single cells from five human lung adenocarcinoma cell lines & GSM3618014 & $3913 \times 11786$ & $63.04$ & 10x Chromium \\
guo~\cite{guo2020singlecell} & Human PBMCs from 2 COVID-19 patients. & GSE150861 & $18177 \times	18538$ & $95.75$ & 10x Chromium \\
itc~\cite{gutierrez-arcelus2019lymphocyte} & Human innate T cells (ITCs). & GSE124731 & $2005 \times 13260$ & $93.05$ & 10x Chromium \\
hca\_10x\_tissue & Bone marrow cells from sample MantonBM6 & HCA\tnote{d} & $6515 \times 18203$ & $90.15$ & 10x Chromium \\
cellmix1~\cite{tian2019benchmarking,tian2018scpipe} & Pseudo cells from nine cell mixtures & GSE118767 & $263 \times 11798$ & $81.41$ & CEL-seq2 \\
rnamix\_celseq2~\cite{tian2018scpipe,tian2019benchmarking} & Pseudo cells from RNA mixtures & GSM3305230 & $340 \times 14804$ & $52.07$ & CEL-seq2 \\
sc\_celseq2~\cite{tian2019benchmarking,tian2018scpipe} & Single cells from three human lung adenocarcinoma cell lines & GSM3336845 & $273 \times 22014$ & $67.82$ & CEL-seq2 \\
sc\_celseq2\_5cl\_p1 \cite{tian2018scpipe,tian2019benchmarking} & Single cells from five human lung adenocarcinoma cell lines & GSM3618022 & $291 \times 15564$	& $61.43$ & CEL-seq2 \\
hcc~\cite{zheng2017landscape} & T cells from hepatocellular carcinoma (HCC). & GSE98638 & $5035 \times 21576$ & $83.94$ & SMART-seq2 \\
petropoulos~\cite{petropoulos2016singlecell} & Human preimplantation embryo scRNA-seq dataset across early developmental stages & E-MTAB-3929 & $1517 \times 23583$ & $57.91$ & SMART-seq2 \\
chu\_cell\_type~\cite{chu2016singlecell} & Human embryonic stem cell scRNA-seq dataset with defined cell states. & GSE75748 & $1018 \times 17559$ & $45.24$ & SMART-seq \\
chu\_time\_course \cite{chu2016singlecell} & Human embryonic stem cell scRNA-seq dataset following differentiation over time. & GSE75748 & $758 \times 16863$ & $48.45$ & SMART-seq \\
chen~\cite{chen2017singlecell} & Mus musculus scRNA-seq dataset of adult mouse hypothalamus revealing diverse neuronal and non-neuronal cell types & GSE87544 & $13891 \times 17623$ & $91.92$ & Drop-seq \\
romanov~\cite{romanov2017molecular} & Mus musculus brain scRNA-seq dataset profiling hypothalamic neuronal cell types. & GSE74672 & $2863 \times 18496$ & $84.06$ & Drop-seq \\
sc\_dropseq~\cite{tian2019benchmarking,tian2018scpipe} & Single cells from three human lung adenocarcinoma cell lines & GSM3336849 & $224 \times 15113$ & $62.13$ & Drop-seq \\
usoskin~\cite{usoskin2015unbiased} & Mouse sensory neuron scRNA-seq dataset profiling dorsal root ganglion cell types. & GSE59739 & $845 \times 16943$ & $80.15$ & STRT-seq \\
zeisel~\cite{zeisel2015cell} & Mouse brain scRNA-seq dataset defining major neuronal and glial cell types. & GSE60361 & $2998 \times 18869$ & $80.17$ & STRT-seq \\
baron~\cite{baron2016singlecell} & Human pancreas scRNA-seq dataset profiling endocrine and exocrine cell types. & GSM2230757 & $1918 \times 14708$ & $87.03$ & inDrop \\
encode\_fluidigm\_5cl \cite{li2017reference} & Single cells from five cell lines & GSE81861 & $360 \times 36092$ & $63.41$ & Fluidigm C1 \\
bladder~\cite{han2018mapping} & Mus musculus bladder scRNA-seq dataset from the Mouse Cell Atlas profiling cell types across bladder tissue. & Figshare\tnote{e} & $2668 \times 15901$ & $93.82$ & Microwell-seq \\
rnamix\_sortseq~\cite{tian2018scpipe,tian2019benchmarking} & Pseudo cells from RNA mixtures & GSM3305231 & $296 \times 15571$ & $56.49$ & Sort-seq \\
\bottomrule
\multicolumn{6}{r}{\textit{Continued on next page}} \\
\end{tabular*}
\begin{tablenotes}
\scriptsize
\item[a] \url{https://www.10xgenomics.com/datasets/jurkat-cells-1-standard-1-1-0}
\item[b] \url{https://www.10xgenomics.com/datasets/293-t-cells-1-standard-1-1-0}
\item[c] \url{https://www.10xgenomics.com/datasets/4-k-pbm-cs-from-a-healthy-donor-2-standard-2-1-0}
\item[d] \url{https://explore.data.humancellatlas.org/projects/cc95ff89-2e68-4a08-a234-480eca21ce79}
\item[e] \url{https://figshare.com/s/865e694ad06d5857db4b}
\end{tablenotes}
\end{threeparttable}
\end{table*}

\begin{table*}[t]
\centering
\fontsize{6.5pt}{7.8pt}\selectfont
\renewcommand{\arraystretch}{1.0}
\setlength{\tabcolsep}{3pt}
\renewcommand*{\multicitedelim}{\addcomma\addnbspace}
\begin{threeparttable}
\begin{tabular*}{\textwidth}{@{\extracolsep{\fill}}>{\raggedright\arraybackslash}p{2.9cm} >{\raggedright\arraybackslash}p{4.05cm} >{\raggedright\arraybackslash}p{52.5pt} >{\raggedleft\arraybackslash}p{61pt} >{\raggedleft\arraybackslash}p{40pt} >{\raggedright\arraybackslash}p{52pt}@{}}
\multicolumn{6}{r}{\textit{Continued from previous page}} \\
\toprule
\textbf{Dataset} & \textbf{Description} & \textbf{Source} & \makecell[tr]{\textbf{Size}\\\textbf{($\mathrm{\mathbf{Cells}} \times \mathrm{\mathbf{Genes}}$)}} & \makecell[tr]{\textbf{Sparsity}\\\textbf{Rates (\%)}} & \textbf{Protocol} \\
simulated\_1 & Splatter simulation with 5 clusters (dropout rate $30.79\,\%$) & - & $2000 \times 600$ & $40.24$\tnote{f} & - \\
simulated\_2 & Splatter simulation with 5 clusters (dropout rate $50.62\,\%$) & - & $2000 \times 600$ & $57.36$\tnote{f} & - \\
simulated\_3 & Splatter simulation with 5 clusters (dropout rate $70.12\,\%$) & - & $2000 \times 600$ & $74.20$\tnote{f} & - \\
simulated\_4 & Splatter simulation with 5 clusters (dropout rate $89.61\,\%$) & - & $2000 \times 600$ & $91.03$\tnote{f} & - \\
\bottomrule
\end{tabular*}
\begin{tablenotes}
\scriptsize
\item[f] Calculated after dropout. The sparsity rate before dropout is $13.65\,\%$ for all 4 simulated datasets.
\end{tablenotes}
\end{threeparttable}
\end{table*}

%% file: tables/nmi.tex
\begin{sidewaystable*}[p]
\renewcommand{\arraystretch}{1.6}
\setlength{\tabcolsep}{1pt}
\caption{NMI of cell clustering based on the imputed and ground truth expression values. The bold values in each row represent the best performance methods. $^{\dagger}$Run with reduced settings (see \cref{subsec:method_experimental_setup}).}\label{tab:nmi}
\fontsize{6pt}{7.5pt}\selectfont
\begin{tabular*}{\textheight}{@{}l@{\extracolsep{\fill}}*{17}{c}@{}}
\toprule
Dataset & Masked & PbImpute & scImpute & SAVER & AcImpute & scTsI & MAGIC & PBLR & scLRTC & WEDGE & scIDPMs$^{\dagger}$ & scSTD & scMultiGAN & scIGANs & scGNN & CPARI & Bubble \\
\midrule
ad\_case & 0.916 & 0.459 & 0.865 & 0.832 & 0.690 & 0.590 & 0.802 & 0.758 & 0.905 & 0.864 & 0.496 & 0.797 & 0.551 & 0.730 & \textbf{0.907} & 0.892 & 0.771 \\
jurkat & 0.287 & 0.182 & 0.346 & 0.320 & 0.244 & 0.183 & 0.300 & 0.048 & 0.294 & 0.440 & 0.091 & 0.295 & 0.235 & 0.099 & 0.333 & \textbf{0.503} & 0.296 \\
293t & 0.244 & 0.141 & 0.315 & 0.202 & 0.123 & 0.032 & 0.194 & 0.039 & 0.248 & 0.372 & 0.051 & 0.219 & 0.107 & 0.051 & 0.263 & \textbf{0.541} & 0.215 \\
pbmc4k & 0.844 & 0.546 & 0.819 & 0.743 & 0.706 & 0.416 & 0.652 & 0.572 & \textbf{0.844} & 0.792 & 0.688 & 0.656 & 0.613 & 0.496 & 0.842 & 0.830 & 0.678 \\
sc\_10x & 0.882 & 0.716 & 0.982 & 0.882 & 0.980 & \textbf{1.000} & 0.751 & 0.192 & 0.882 & 0.961 & 0.704 & 0.743 & \textbf{1.000} & 0.882 & 0.980 & 0.980 & 0.949 \\
sc\_10x\_5cl & 0.948 & 0.694 & 0.803 & 0.859 & 0.824 & 0.622 & 0.719 & 0.157 & \textbf{0.906} & 0.788 & 0.843 & 0.697 & 0.783 & 0.708 & 0.824 & 0.881 & 0.841 \\
guo & 0.843 & 0.493 & 0.824 & 0.814 & 0.586 & 0.633 & 0.717 & 0.708 & \textbf{0.848} & 0.820 & 0.404 & 0.795 & 0.592 & 0.831 & 0.819 & 0.830 & 0.662 \\
itc & 0.365 & 0.373 & 0.423 & 0.395 & 0.276 & 0.030 & 0.306 & 0.039 & 0.184 & 0.221 & 0.048 & 0.315 & 0.424 & 0.102 & \textbf{0.445} & 0.398 & 0.338 \\
hca\_10x\_tissue & 0.856 & 0.591 & 0.793 & 0.787 & 0.680 & 0.269 & 0.702 & 0.597 & 0.807 & 0.827 & 0.421 & 0.751 & 0.565 & 0.575 & \textbf{0.873} & 0.833 & 0.722 \\
cellmix1 & 0.121 & 0.048 & 0.175 & 0.242 & 0.036 & 0.077 & \textbf{0.292} & 0.059 & 0.098 & 0.017 & 0.040 & 0.010 & 0.045 & 0.044 & 0.260 & 0.175 & 0.191 \\
rnamix\_celseq2 & 0.371 & 0.191 & \textbf{0.888} & 0.462 & 0.700 & 0.113 & 0.636 & 0.055 & 0.371 & 0.512 & 0.085 & 0.680 & 0.024 & 0.082 & 0.589 & 0.370 & 0.429 \\
sc\_celseq2 & 0.817 & 0.005 & 0.567 & 0.538 & 0.579 & 0.350 & 0.536 & 0.098 & \textbf{0.835} & 0.402 & 0.175 & 0.520 & 0.232 & 0.151 & 0.656 & 0.464 & 0.722 \\
sc\_celseq2\_5cl\_p1 & 0.576 & 0.252 & \textbf{0.670} & 0.562 & 0.612 & 0.539 & 0.595 & 0.062 & 0.507 & 0.589 & 0.081 & 0.562 & 0.438 & 0.008 & 0.475 & 0.611 & 0.454 \\
hcc & 0.820 & 0.369 & 0.638 & 0.804 & 0.482 & 0.449 & 0.729 & 0.023 & \textbf{0.858} & 0.660 & 0.425 & 0.670 & 0.507 & 0.162 & 0.731 & 0.845 & 0.710 \\
petropoulos & 0.861 & 0.406 & 0.850 & 0.905 & 0.839 & 0.729 & 0.759 & 0.056 & 0.892 & 0.813 & 0.490 & 0.695 & 0.805 & 0.559 & 0.816 & \textbf{0.907} & 0.854 \\
chu\_cell\_type & 0.932 & 0.510 & 0.881 & 0.945 & 0.906 & 0.898 & 0.853 & 0.048 & 0.932 & 0.858 & 0.920 & 0.769 & 0.933 & 0.889 & 0.945 & 0.885 & \textbf{1.000} \\
chu\_time\_course & 0.891 & 0.523 & 0.958 & 0.874 & 0.912 & 0.855 & 0.825 & 0.046 & 0.891 & 0.804 & 0.543 & 0.814 & 0.919 & \textbf{0.976} & 0.912 & 0.933 & 0.939 \\
chen & 0.932 & 0.604 & 0.849 & 0.889 & 0.747 & 0.722 & 0.827 & 0.716 & \textbf{0.930} & 0.857 & 0.669 & 0.825 & 0.562 & 0.842 & 0.898 & 0.902 & 0.829 \\
romanov & 0.818 & 0.496 & 0.647 & 0.751 & 0.615 & 0.549 & 0.785 & 0.200 & 0.792 & 0.766 & 0.669 & 0.731 & 0.534 & 0.435 & 0.820 & \textbf{0.836} & 0.764 \\
sc\_dropseq & 0.702 & 0.324 & 0.658 & 0.850 & 0.735 & 0.561 & 0.850 & 0.022 & 0.561 & 0.853 & 0.684 & 0.853 & 0.607 & 0.332 & 0.850 & 0.803 & \textbf{0.876} \\
usoskin & 0.822 & 0.444 & 0.697 & 0.773 & 0.260 & 0.502 & 0.591 & 0.075 & \textbf{0.822} & 0.326 & 0.254 & 0.736 & 0.584 & 0.263 & 0.757 & 0.821 & 0.602 \\
zeisel & 0.857 & 0.648 & 0.790 & 0.816 & 0.666 & 0.652 & 0.774 & 0.224 & 0.857 & 0.831 & 0.584 & 0.756 & 0.544 & 0.463 & 0.863 & \textbf{0.890} & 0.821 \\
baron & 0.893 & 0.486 & 0.656 & 0.748 & 0.644 & 0.147 & 0.696 & 0.137 & \textbf{0.875} & 0.741 & 0.397 & 0.743 & 0.358 & 0.628 & 0.830 & 0.853 & 0.720 \\
encode\_fluidigm\_5cl & 0.730 & 0.323 & 0.901 & 0.730 & 0.740 & 0.587 & 0.876 & 0.148 & 0.730 & 0.747 & 0.732 & 0.695 & \textbf{0.924} & 0.825 & 0.771 & 0.883 & 0.922 \\
bladder & 0.863 & 0.517 & 0.742 & 0.798 & 0.718 & 0.636 & 0.822 & 0.790 & 0.810 & 0.603 & 0.728 & 0.778 & 0.388 & 0.626 & \textbf{0.831} & 0.781 & 0.725 \\
rnamix\_sortseq & 0.303 & 0.076 & 0.357 & 0.262 & 0.307 & 0.143 & 0.291 & 0.050 & 0.303 & 0.131 & 0.016 & 0.410 & 0.052 & 0.046 & 0.281 & \textbf{0.440} & 0.431 \\
simulated\_1 & 0.671 & 0.332 & 0.921 & 0.749 & 0.944 & \textbf{1.000} & 0.793 & 0.047 & 0.896 & \textbf{1.000} & 0.024 & 0.784 & 0.692 & 0.774 & 0.680 & 0.983 & 0.983 \\
simulated\_2 & 0.166 & 0.077 & 0.372 & 0.235 & 0.295 & 0.958 & 0.602 & 0.054 & 0.161 & \textbf{1.000} & 0.035 & 0.574 & 0.620 & 0.255 & 0.310 & 0.692 & 0.690 \\
simulated\_3 & 0.052 & 0.106 & 0.046 & 0.078 & 0.047 & 0.236 & 0.231 & 0.042 & 0.091 & \textbf{1.000} & 0.024 & 0.085 & 0.132 & 0.128 & 0.069 & 0.134 & 0.021 \\
simulated\_4 & 0.021 & 0.047 & 0.015 & 0.030 & 0.059 & 0.019 & 0.049 & 0.022 & 0.022 & \textbf{0.078} & 0.023 & 0.027 & 0.033 & 0.044 & 0.038 & 0.030 & 0.013 \\
\bottomrule
\end{tabular*}%
\end{sidewaystable*}

%% file: tables/purity.tex
\begin{sidewaystable*}[p]
\renewcommand{\arraystretch}{1.6}
\setlength{\tabcolsep}{1pt}
\caption{Purity of cell clustering based on the imputed and ground truth expression values. The bold values in each row represent the best performance methods. $^{\dagger}$Run with reduced settings (see \cref{subsec:method_experimental_setup}).}\label{tab:purity}
\fontsize{6pt}{7.5pt}\selectfont
\begin{tabular*}{\textheight}{@{}l@{\extracolsep{\fill}}*{17}{c}@{}}
\toprule
Dataset & Masked & PbImpute & scImpute & SAVER & AcImpute & scTsI & MAGIC & PBLR & scLRTC & WEDGE & scIDPMs$^{\dagger}$ & scSTD & scMultiGAN & scIGANs & scGNN & CPARI & Bubble \\
\midrule
ad\_case & 0.938 & 0.621 & 0.926 & 0.907 & 0.798 & 0.727 & 0.938 & 0.860 & 0.929 & 0.922 & 0.547 & 0.916 & 0.687 & 0.811 & 0.958 & \textbf{0.963} & 0.828 \\
jurkat & 0.451 & 0.378 & 0.452 & 0.448 & 0.412 & 0.346 & 0.434 & 0.227 & 0.468 & 0.576 & 0.259 & 0.418 & 0.389 & 0.259 & 0.465 & \textbf{0.614} & 0.452 \\
293t & 0.443 & 0.334 & 0.374 & 0.357 & 0.367 & 0.245 & 0.384 & 0.240 & 0.437 & 0.569 & 0.273 & 0.390 & 0.283 & 0.266 & 0.414 & \textbf{0.744} & 0.402 \\
pbmc4k & 0.917 & 0.773 & 0.929 & 0.919 & 0.883 & 0.650 & 0.882 & 0.770 & 0.919 & 0.923 & 0.800 & 0.903 & 0.847 & 0.737 & 0.932 & \textbf{0.941} & 0.836 \\
sc\_10x & 0.856 & 0.856 & 0.994 & 0.856 & 0.994 & \textbf{1.000} & 0.983 & 0.569 & 0.856 & 0.989 & 0.796 & 0.983 & \textbf{1.000} & 0.856 & 0.994 & 0.994 & 0.983 \\
sc\_10x\_5cl & 0.980 & 0.926 & \textbf{0.980} & \textbf{0.980} & \textbf{0.980} & 0.890 & \textbf{0.980} & 0.484 & 0.839 & 0.978 & 0.969 & 0.974 & 0.966 & 0.925 & \textbf{0.980} & \textbf{0.980} & \textbf{0.980} \\
guo & 0.806 & 0.583 & 0.857 & 0.829 & 0.632 & 0.662 & 0.796 & 0.702 & 0.817 & 0.856 & 0.431 & \textbf{0.866} & 0.550 & 0.800 & 0.852 & 0.821 & 0.666 \\
itc & 0.759 & 0.808 & 0.806 & 0.808 & 0.692 & 0.443 & 0.769 & 0.448 & 0.657 & 0.694 & 0.505 & 0.781 & 0.791 & 0.520 & \textbf{0.821} & 0.813 & 0.751 \\
hca\_10x\_tissue & 0.847 & 0.671 & 0.810 & 0.816 & 0.778 & 0.396 & 0.773 & 0.598 & 0.769 & 0.832 & 0.498 & 0.861 & 0.570 & 0.606 & \textbf{0.892} & 0.844 & 0.738 \\
cellmix1 & 0.698 & 0.604 & 0.717 & \textbf{0.811} & 0.623 & 0.679 & \textbf{0.811} & 0.642 & 0.698 & 0.585 & 0.623 & 0.566 & 0.642 & 0.623 & \textbf{0.811} & 0.755 & 0.774 \\
rnamix\_celseq2 & 0.783 & 0.551 & \textbf{0.971} & 0.739 & 0.928 & 0.435 & 0.870 & 0.478 & 0.783 & 0.783 & 0.522 & 0.928 & 0.420 & 0.522 & 0.870 & 0.696 & 0.739 \\
sc\_celseq2 & 0.891 & 0.455 & 0.818 & 0.800 & 0.818 & 0.655 & 0.764 & 0.527 & \textbf{0.891} & 0.673 & 0.600 & 0.818 & 0.545 & 0.455 & 0.855 & 0.764 & 0.873 \\
sc\_celseq2\_5cl\_p1 & 0.898 & 0.678 & \textbf{0.932} & 0.881 & 0.915 & 0.881 & 0.898 & 0.492 & 0.864 & 0.898 & 0.492 & 0.881 & 0.780 & 0.441 & 0.864 & 0.898 & 0.847 \\
hcc & 0.885 & 0.460 & 0.725 & 0.898 & 0.596 & 0.535 & 0.912 & 0.205 & \textbf{0.935} & 0.759 & 0.427 & 0.810 & 0.611 & 0.307 & 0.808 & 0.911 & 0.804 \\
petropoulos & 0.938 & 0.498 & 0.928 & \textbf{0.954} & 0.869 & 0.800 & 0.898 & 0.279 & 0.948 & 0.902 & 0.508 & 0.856 & 0.882 & 0.662 & 0.846 & 0.921 & 0.931 \\
chu\_cell\_type & 0.990 & 0.649 & 0.995 & 0.995 & 0.966 & 0.985 & 0.980 & 0.385 & 0.990 & 0.985 & 0.985 & 0.917 & 0.990 & 0.927 & 0.995 & 0.893 & \textbf{1.000} \\
chu\_time\_course & 0.967 & 0.725 & 0.987 & 0.961 & 0.974 & 0.954 & 0.974 & 0.431 & 0.967 & 0.967 & 0.758 & 0.980 & 0.974 & \textbf{0.993} & 0.974 & 0.980 & 0.980 \\
chen & 0.905 & 0.714 & 0.912 & 0.891 & 0.808 & 0.762 & \textbf{0.932} & 0.680 & 0.906 & 0.869 & 0.608 & 0.899 & 0.639 & 0.852 & 0.884 & \textbf{0.932} & 0.830 \\
romanov & 0.846 & 0.595 & 0.672 & 0.770 & 0.688 & 0.637 & 0.845 & 0.312 & 0.838 & 0.813 & 0.642 & 0.836 & 0.532 & 0.532 & \textbf{0.918} & 0.860 & 0.764 \\
sc\_dropseq & 0.891 & 0.609 & 0.891 & \textbf{0.957} & 0.891 & 0.826 & \textbf{0.957} & 0.435 & 0.826 & \textbf{0.957} & 0.913 & \textbf{0.957} & 0.848 & 0.674 & \textbf{0.957} & 0.935 & \textbf{0.957} \\
usoskin & 0.900 & 0.624 & 0.835 & 0.882 & 0.471 & 0.659 & 0.776 & 0.347 & 0.900 & 0.506 & 0.488 & 0.894 & 0.694 & 0.476 & 0.894 & \textbf{0.924} & 0.718 \\
zeisel & 0.889 & 0.745 & 0.837 & 0.820 & 0.757 & 0.719 & 0.872 & 0.323 & 0.889 & 0.855 & 0.572 & 0.857 & 0.614 & 0.527 & 0.912 & \textbf{0.918} & 0.860 \\
baron & 0.961 & 0.660 & 0.774 & 0.847 & 0.839 & 0.449 & 0.849 & 0.418 & \textbf{0.953} & 0.925 & 0.629 & 0.930 & 0.590 & 0.810 & \textbf{0.953} & 0.948 & 0.857 \\
encode\_fluidigm\_5cl & 0.808 & 0.493 & 0.959 & 0.808 & 0.877 & 0.726 & 0.945 & 0.466 & 0.808 & 0.890 & 0.836 & 0.795 & \textbf{0.973} & 0.863 & 0.836 & 0.959 & \textbf{0.973} \\
bladder & 0.935 & 0.636 & 0.779 & 0.899 & 0.858 & 0.809 & \textbf{0.927} & 0.778 & 0.877 & 0.716 & 0.774 & 0.892 & 0.533 & 0.753 & 0.847 & 0.815 & 0.793 \\
rnamix\_sortseq & 0.650 & 0.417 & 0.667 & 0.600 & 0.633 & 0.450 & 0.650 & 0.367 & 0.650 & 0.500 & 0.383 & 0.700 & 0.417 & 0.383 & 0.550 & 0.717 & \textbf{0.733} \\
simulated\_1 & 0.870 & 0.545 & 0.975 & 0.905 & 0.983 & \textbf{1.000} & 0.927 & 0.333 & 0.968 & \textbf{1.000} & 0.268 & 0.960 & 0.897 & 0.917 & 0.877 & 0.995 & 0.995 \\
simulated\_2 & 0.435 & 0.357 & 0.693 & 0.537 & 0.605 & 0.988 & 0.845 & 0.333 & 0.453 & \textbf{1.000} & 0.300 & 0.835 & 0.848 & 0.560 & 0.588 & 0.873 & 0.873 \\
simulated\_3 & 0.345 & 0.388 & 0.312 & 0.388 & 0.323 & 0.487 & 0.537 & 0.335 & 0.398 & \textbf{1.000} & 0.292 & 0.350 & 0.438 & 0.388 & 0.347 & 0.448 & 0.268 \\
simulated\_4 & 0.265 & 0.338 & 0.258 & 0.297 & 0.338 & 0.263 & 0.315 & 0.270 & 0.273 & \textbf{0.357} & 0.270 & 0.290 & 0.297 & 0.318 & 0.302 & 0.297 & 0.265 \\
\bottomrule
\end{tabular*}%
\end{sidewaystable*}

%% file: tables/performance_summary.tex
\begin{table*}[tb]
\centering
\newcommand{\pbest}{\cellcolor[HTML]{C6EFCE}{\color[HTML]{006100}$\bigstar$}}
\newcommand{\pmid}{\cellcolor[HTML]{FFEB9C}{\color[HTML]{9C5700}$\triangle$}}
\newcommand{\pworst}{\cellcolor[HTML]{FFC7CE}{\color[HTML]{9C0006}$\times$}}
\renewcommand{\arraystretch}{1.6}
\setlength{\aboverulesep}{0pt}
\setlength{\belowrulesep}{0pt}
\caption{Summary of the performance of imputation methods. {\color[HTML]{006100}$\bigstar$}, {\color[HTML]{9C5700}$\triangle$}, and {\color[HTML]{9C0006}$\times$} indicate best, moderate, and worst performance, respectively. $^{\dagger}$Run with reduced settings (see \cref{subsec:method_experimental_setup}).}
\label{tab:performance_summary}
\fontsize{8pt}{10pt}\selectfont
\setlength{\tabcolsep}{2pt}
\begin{tabular}{ll cccccc}
\toprule
\textbf{Category} & \textbf{Method}
  & \makecell{\textbf{Numerical gene}\\\textbf{expression recovery}}
  & \makecell{\textbf{Cell}\\\textbf{clustering}}
  & \makecell{\textbf{DE}\\\textbf{analysis}}
  & \makecell{\textbf{Marker gene}\\\textbf{analysis}}
  & \makecell{\textbf{Trajectory}\\\textbf{analysis}}
  & \makecell{\textbf{Cell type}\\\textbf{annotation}} \\
\midrule
\multirow{3}{*}{\makecell[l]{Model-based}}
& PbImpute            & \pmid   & \pmid   & \pmid   & \pworst & \pworst & \pmid   \\
& scImpute            & \pmid   & \pmid   & \pmid   & \pbest  & \pbest  & \pmid   \\
& SAVER               & \pmid   & \pmid   & \pbest  & \pbest  & \pbest  & \pmid   \\
\midrule
\multirow{3}{*}{\makecell[l]{Smoothing-\\based}}
& AcImpute            & \pmid   & \pmid   & \pmid   & \pmid   & \pmid   & \pmid   \\
& scTsI               & \pmid   & \pmid   & \pmid   & \pmid   & \pworst & \pmid   \\
& MAGIC               & \pmid   & \pbest  & \pmid   & \pmid   & \pmid   & \pbest  \\
\midrule
\multirow{3}{*}{\makecell[l]{Low-rank\\matrix-based}}
& PBLR                & \pmid   & \pworst & \pworst & \pworst & \pworst & \pmid   \\
& scLRTC              & \pmid   & \pbest  & \pworst & \pbest  & \pbest  & \pmid   \\
& WEDGE               & \pbest  & \pmid   & \pmid   & \pmid   & \pmid   & \pmid   \\
\midrule
\multirow{2}{*}{\makecell[l]{Diffusion-\\based}}
& scIDPMs$^{\dagger}$ & \pworst & \pworst & \pworst & \pworst & \pworst & \pworst \\
& scSTD               & \pmid   & \pmid   & \pmid   & \pmid   & \pmid   & \pmid   \\
\midrule
\multirow{2}{*}{\makecell[l]{GAN-based}}
& scMultiGAN          & \pmid   & \pmid   & \pworst & \pmid   & \pmid   & \pmid   \\
& scIGANs             & \pworst & \pmid   & \pworst & \pmid   & \pworst & \pmid   \\
\midrule
\makecell[l]{GNN-based}
& scGNN               & \pmid   & \pmid   & \pmid   & \pmid   & \pmid   & \pmid   \\
\midrule
\multirow{2}{*}{\makecell[l]{AE-based}}
& CPARI               & \pmid   & \pmid   & \pmid   & \pbest  & \pbest  & \pmid   \\
& Bubble              & \pmid   & \pmid   & \pmid   & \pmid   & \pbest  & \pmid   \\
\bottomrule
\end{tabular}
\end{table*}